\documentclass[lettersize,journal,compsoc]{IEEEtran}

\usepackage{amsmath,amsfonts}
\usepackage{algorithmic}
\usepackage{array}
\usepackage{textcomp}
\usepackage{stfloats}
\usepackage{url}
\usepackage{verbatim}
\usepackage{graphicx}
\def\BibTeX{{\rm B\kern-.05em{\sc i\kern-.025em b}\kern-.08em
    T\kern-.1667em\lower.7ex\hbox{E}\kern-.125emX}}
\usepackage{balance}

\makeatletter
\DeclareRobustCommand\onedot{\futurelet\@let@token\@onedot}
\def\@onedot{\ifx\@let@token.\else.\null\fi\xspace}

\usepackage[hidelinks]{hyperref}
\usepackage{enumitem}
\usepackage[svgnames]{xcolor}
\usepackage{amsmath}
\usepackage{comment}
\usepackage[frozencache,cachedir=minted-cache]{minted} 
\usepackage{changepage}
\usepackage[font=normalsize,labelfont=sf,textfont=sf]{subcaption}

\usepackage{tikz}
\newcommand\node[1]{\tikz[baseline=(x.base)]{
  \node[draw,rounded corners] (x) {#1};
}}

\setminted[javascript]{ %
    breaklines,
    breakafter=(, 
    linenos=true,             
    numbersep=2pt,
    autogobble=true,          
    frame=lines,
    framesep=2mm,
    framerule=0.1pt,
    fontsize=\small
}

\newif\ifnotes
\notesfalse

\newcommand{\revised}[1]{\ifnotes{\leavevmode\color{blue} #1}\else{#1}\fi}

\newcommand{\eg}[0]{e.g.,}

\newcommand{\model}[0]{Pyxis}

\begin{document}

\title{What Do We Mean When We Say ``Insight''?\\ A \revised{Formal} Synthesis of Existing Theory}

\author{Leilani Battle and Alvitta Ottley\thanks{Leilani Battle is with the University of Washington, Seattle (e-mail: leibatt@uw.edu). Alvitta Ottley is with Washington University in St. Louis (e-mail: alvitta@wustl.edu).}\vspace{-4mm}}


\IEEEtitleabstractindextext{%
\begin{abstract}
Researchers have derived many theoretical models for specifying users’ insights as they interact with a visualization system. These representations are essential for understanding the insight discovery process, such as when inferring user interaction patterns that lead to insight or assessing the rigor of reported insights. However, theoretical models can be difficult to apply to existing tools and user studies, often due to discrepancies in how insight and its constituent parts are defined. This paper calls attention to the consistent structures that recur across the visualization literature and describes how they connect multiple theoretical representations of insight. We synthesize a unified formalism for insights using these structures, enabling a wider audience of researchers and developers to adopt the corresponding models. Through a series of theoretical case studies, we use our formalism to compare and contrast existing theories, revealing interesting research challenges in reasoning about a user's domain knowledge and leveraging synergistic approaches in data mining and data management research.

\end{abstract}

}

\maketitle

\IEEEdisplaynontitleabstractindextext

%
\IEEEpeerreviewmaketitle

\ifCLASSOPTIONcompsoc
\IEEEraisesectionheading{\section{Introduction}\label{sec:introduction}}
\else
\section{Introduction}
\label{sec:introduction}
\fi


\vspace{-1mm}

\IEEEPARstart{B}{en} Shneiderman famously said, ``the purpose of visualization is \textit{insight}, not pictures,'' underscoring a long-held conviction in the visualization community that the goal of the visualization tool is to enhance the user's understanding of the underlying data~\cite{hullman2019purpose}. To this end, numerous scholars have aimed to understand the insight discovery process, yielding a wide range of \emph{theories} defining an insight's structure, properties, and efficacy.  These theories are critical in advancing insight-based research, producing models, processes, and metrics that researchers can adopt to derive meaning from their empirical observations. For example, researchers have used theory to infer a user’s objective through recurring patterns in her logged interactions~\cite{battle2016dynamic,monadjemi_competing_2020,gathani2022grammar,pohl2012analysing}, or to understand which of the user’s utterances correspond to high-quality insights~\cite{north_toward_2006,zgraggen_investigating_2018}.

However, the distinguishing characteristics of individual theories, the relationships between them, and their appropriateness for a given evaluation scenario are still unclear.
For example, \revised{what does each theory mean by saying ``insight''?}
When theories disagree, what are the benefits and drawbacks of adopting one over another? Finally, what gaps exist in the literature that new theories could fill? 

In pursuit of these questions, we present a \emph{theory-driven exploration} of existing work. First, we summarize existing arguments on the role and structure of \revised{insight}.
Then, we synthesize the core building blocks of insight based on observed overlaps in existing theories and contribute a formal specification for each building block. Finally, we compare the benefits and costs of several established theories of insight through case studies based on how they align with or deviate from our theorized building blocks.

We observe that researchers often differ in their terminology for theoretical concepts~\cite{battle_characterizing_2019,gathani2022grammar,rind_task_2016}, making rote interpretation of the original papers a challenging strategy for making meaningful comparisons. Inspired by research in visualization specification~\cite{satyanarayan2017vegalite,McNutt2021Templates}, we adopt a specification-based approach, where we use language-based specifications of the core building blocks of insight (as synthesized from the literature) to implement the structures implied in the original papers. Then, we analyze the resulting structures to identify interesting agreements and/or disagreements between existing definitions. Through these specifications, we place each theory on equal ground and make precise theoretical comparisons. Furthermore, we publish each implemented case study as a self-contained, executable program that can be evaluated or reused by the visualization community in the future\footnote{\url{https://osf.io/t9e63/?view_only=9857ef52c3334739b0001dd6d7cf324c}}.

Based on our findings, we highlight several exciting research opportunities, such as formally defining and capturing domain knowledge, sharing insight data alongside system interaction logs, and broader use of innovations in data mining and data management across the visualization community.
We discuss these challenges further in \autoref{sec:discussion}.

In summary, this paper makes \revised{four} contributions:
\begin{itemize}
\item We present an in-depth \textbf{\revised{review of existing insight definitions} and connect them} with related concepts in insight discovery, such as analysis \textit{tasks}.
\item Based on observed agreements in the literature, we \textbf{synthesize the core building blocks of insights} and contribute formal specifications of these building blocks.
\item \revised{We introduce \textbf{four use cases for applying our formalism} to a range of insight-based theories and studies.}
\item We \textbf{\revised{discuss} open challenges in insight- and theory-based visualization research} motivated by this work.
\end{itemize}
\section{How are Insights \revised{Currently Defined?}}
\label{sec:background}


In this section, we review the visualization literature to understand how insights are conceptualized and to identify \emph{agreements} and \emph{disagreements} among existing theories.
\revised{Note that this work summarizes a prior literature review~\cite{battle2023exactly}.}


\subsection{Review Process} 
\revised{We conducted keyword searches for
``visualization insight''
and
``visualization task''
in Google Scholar; we reviewed the proceedings of VIS and EuroVis from 2013 to 2023; and we noted relevant papers with ``insight'' in the title or abstract.}
This generated an initial list of 125 papers.
\revised{Then, we filtered for only papers 
that explicitly investigate how to define, analyze, or \revised{discover} insights}, e.g., ``But what, exactly, is insight? How can it be measured and evaluated?''~\cite{north_comparison_2011}.
For each relevant paper, we reviewed its list of references to identify papers we may have missed. 
\revised{To explore potential relationships between insights and existing task theory, we also considered papers cited by our initial list that explicitly defined visualization objectives, tasks, or reasoning models.}
These steps yielded a list of 38 papers. With feedback from colleagues/reviewers, we extended it to include their suggestions, producing a final list of \revised{42} papers.
The two authors collaboratively analyzed how insight was defined in each paper, focusing on high-level themes as well as key characteristics of insights. We cite synergistic ideas when relevant, e.g., knowledge graphs and visualization recommendations.

\vspace{-1mm}
\subsection{Categories of Insight}
\label{sec:background:insight:categories}


The prior work details several high-level categories of insights, \revised{distinguishing} \textit{instantaneous sparks} (``aha'' moments) and long-term \textit{knowledge building}. For example,  Chang et al.~\cite{chang_defining_2009} distinguish between a ``knowledge-building insight,'' or information that \revised{extends} a user's existing knowledge structures, and ``spontaneous insight,'' or a ``eureka'' moment that reorganizes 
\revised{loosely} related knowledge. 
\revised{We focus on} knowledge-building insights in this paper, but our formalism could be extended to spontaneous insights.

\revised{We also observe categorizations focused} on the \emph{source} of insights, such as the input dataset, social structures of the analyst, or an analyst's domain knowledge.
For example,
Saraiya et al. define four categories of data-driven insights~\cite{saraiya_insight-based_2005,saraiya_evaluation_2004}: overall distributions, patterns, grouping, and detail.
Choe et al.~\cite{choe_characterizing_2015} extend these ideas by providing more granular categorizations, such as distinguishing distributions versus data summaries or correlations versus general trends.
Zgraggen et al.~\cite{zgraggen_investigating_2018} follow a similar structure but focus on categorizing the \revised{ways} people make data comparisons to extract \revised{data-driven insights.}
Note that these categories are not mutually exclusive and \revised{may} co-occur~\cite{smuc_score_2009}.

However, data-driven insights are not the only insights an analyst may uncover. For example, Gotz et al.~\cite{gotz_interactive_2006}, Pousman et al.~\cite{pousman_casual_2007}, Liu and Heer~\cite{liu_effects_2014}, Choe et al.~\cite{choe_characterizing_2015}, and Karer et al.~\cite{karer2021insight} observe that analysts often connect what they see in the data with their own knowledge and experiences, i.e., with \emph{domain knowledge} that exists outside the target dataset. Pousman et al. broaden this view to support insights that may not be purely data-driven, in particular ``awareness insight,'' ``social insight,'' and ``reflective insight''~\cite{pousman_casual_2007}. 

The many variations in how insights are categorized suggest that insights may not be an atomic unit in themselves. Instead, it may be more useful to categorize the \emph{components} of insights, such as the types of knowledge gleaned from data-driven or domain-driven sources~\cite{karer2021insight}. We take this into consideration in our formalism and subsequent analysis.

\vspace{-1mm}
\subsection{The Varying Definitions of Insight}
\label{sec:background:insight:definitions}
\vspace{-1mm}

We also observe
\revised{inconsistent} \emph{definitions} for what constitutes an insight. Are they utterances, statistical correlations, or something more complex? In this section, we summarize the range of definitions proposed in the literature
to identify key components relevant to developing a unified formalism.

\revised{Some definitions assert that \textbf{insights are utterances}. For example, in the context of a user study, Saraiya et al. define insight as ``an individual observation about the data by the participant, a unit of discovery,''~\cite{saraiya_insight-based_2005,saraiya_evaluation_2004} which can include ``any data observation that the user mentions'' during lab studies~\cite{saraiya_insight-based_2005,liu_effects_2014,zgraggen_investigating_2018,zgraggen2017progressive} and self-reported insight diaries from field studies~\cite{saraiya_insight-based_2006} and competition submissions~\cite{plaisant_promoting_2008}.} 

Insights are often categorized by how their calculation supports users' hypotheses, claims, and reflections, pointing to a second definition -- \textbf{insights are data facts}.
For exmaple, Choe et al. propose eight insight classes, where six classes are statistical in nature (``trend,'' ``correlation,'' ``data summary,'' ``distribution,'' ``outlier'' and ``comparison'') and two are adapted from existing taxonomies (``detail''~\cite{saraiya_insight-based_2005,saraiya_insight-based_2006} and ``self-reflection''~\cite{pousman_casual_2007}).
Zgraggen et al. propose five insight classes, all of which are statistical in nature~\cite{zgraggen_investigating_2018}: ``shape,'' ``mean,'' ``variance,'' ``correlation,'' and ``ranking''.
\revised{This consistent grouping suggests that statistical representations, i.e., \emph{data facts} may be a core building block of insights. Chen et al. formalize the relationship between data facts and insights through their Fact Management Framework~\cite{yang_chen_toward_2009}, which provides a theoretical base for defining insights.}

The prior work also suggests that \textbf{insights are hypotheses and/or evidence.} For example, 
Sacha et al. argue that users leverage analysis findings primarily as evidence to support, refute, or generate new hypotheses~\cite{sacha_knowledge_2014}.
To evaluate how study participants perform during open-ended exploration tasks, Gomez et al. label each observed insight from their study as a ``claim,'' i.e., ``a general hypothesis, question, or remark about the data model that is potentially synthesized from multiple observations,'' or as ``evidence,''  such as an observation comprised of ``specific references to data points'' that support the claim~\cite{gomez_insight-_2014}.
\revised{Guo et al.~\cite{guo_case_2016} and Liu and Heer~\cite{liu_effects_2014} adopt a similar evidence- and hypothesis-based framing for insights, respectively.}
\revised{Thus, this definition seems to build on the concept of data facts.}

Finally, the prior work also suggests that \textbf{insights are knowledge links.} In particular, Chang et al. argue that visual analytics research ``considers insight more or less as units of knowledge''~\cite{chang_defining_2009}. Others refine this idea further by defining insights as \emph{links} that connect analysis \emph{findings}, such as visualizations and statistical results~\cite{srinivasan_augmenting_2019, kandogan2018towards}, with user \emph{knowledge} (e.g., ~\cite{saraiya_insight-based_2005,north_toward_2006,gotz_interactive_2006,yi_understanding_2008,amar_knowledge_2005,sacha_knowledge_2014,shrinivasan_connecting_2009,karer2021insight}), such as knowledge synthesized from the current session or earlier sessions (e.g., ~\cite{shrinivasan_supporting_2008,chang_defining_2009,gotz_characterizing_2009,smuc_score_2009,he_characterizing_2020}), or a priori knowledge the user brings to the exploration process (e.g., ~\cite{saraiya_insight-based_2006,gotz_interactive_2006,yi_understanding_2008, pousman_casual_2007,amar_knowledge_2005,green_visual_2008,karer2021insight}).

These links can be implicit, such as when observed through qualitative studies~\cite{saraiya_insight-based_2005,saraiya_insight-based_2006,smuc_score_2009}, or explicit, such as when users apply annotation interactions~\cite{gotz_interactive_2006,gotz_characterizing_2009,willett2011commentspace} or link interactions~\cite{shrinivasan_supporting_2008,gotz_interactive_2006,gotz_characterizing_2009,shrinivasan_connecting_2009,he_characterizing_2020,dou_recovering_2009} to connect system visualization state with concepts recorded in their own digitized notes. Furthermore, insights can be hierarchical and build on one another over time, increasing the complexity of subsequent insights~\cite{saraiya_insight-based_2005,saraiya_insight-based_2006,north_toward_2006,pousman_casual_2007,gotz_characterizing_2009,smuc_score_2009,shrinivasan_connecting_2009,mathisen_insideinsights_2019,green_visual_2008}.

Moreover, Pike et al. argue for more formal semantics for capturing user insights, which can enable visual analysis systems to more effectively process, reason about and even extract new insights~\cite{pike_science_2009}.
Moreover, Smuc et al. argue that insights are better analyzed by explicitly tracking how they build on one another and propose relational insight organizers (or RIOs) to organize and visualize the resulting insight graph~\cite{smuc_score_2009}.
RIOs share similarities with the structures proposed by Gotz et al.~\cite{gotz_interactive_2006}, where user knowledge is also captured as a graph, with high-level concepts and instantiations of these concepts representing nodes in the graph, and links between instances and/or data representing edges in the graph.
Similar graph-based structures have also been suggested by Shrinivasan and van Wijk~\cite{shrinivasan_supporting_2008}, Willett et al.~\cite{willett2011commentspace}, Mathisen et al.~\cite{mathisen_insideinsights_2019}, and He et al.~\cite{he_characterizing_2020}, as well as in the intelligence analysis literature~\cite{toniolo2023human}.

\paragraph*{\textbf{Integrating the Definitions}} 
These definitions may appear distinct. However, a close look at the varying perspectives points to an overarching theme -- \textbf{an insight is a collection of knowledge}. 
Although existing definitions vary in what they emphasize, we find that the components themselves appear to be consistent across definitions, which we categorize as \emph{analytic} and \emph{domain knowledge}. For example, analytic knowledge consistently includes data facts, generalizations, and hypotheses. Domain knowledge includes domain expertise and personal experiences. Awareness of these components enables us to navigate these varied definitions of insight.
Further, our proposed formalism unites these perspectives by distinguishing between these knowledge sources and defining concrete links between them. 

\vspace{-1mm}
\subsection{Scoping Insights}
\label{sec:background:insight:scope}
\vspace{-1mm}

\revised{Insight discovery generally occurs within a certain visual analysis scope~\cite{gotz_characterizing_2009,gathani2022grammar}}
\revised{which is often tightly bound with the definition of tasks~\cite{gathani2022grammar,brehmer_multi-level_2013} or objectives~\cite{rind_task_2016,lam_bridging_2018} in visualization research. For example, many theories categorize the scope of insights that analysts may be looking for.
These theories} often take the form of task \emph{taxonomies} and \emph{typologies}~\cite{gathani2022grammar}, where specific tasks observed in the field or lab studies are
\revised{abstracted into}
task classes, such as ``Find Anomalies''~\cite{amar_low-level_2005}, ``Search/Comparison''~\cite{kang_examining_2012} or ``characterizing data distributions and relationships''~\cite{battle_characterizing_2019}.
Specific to insights, several taxonomies target common insight generation \emph{processes} to predict insight scope, rigor, and complexity~\cite{yi_understanding_2008,guo_case_2016,battle_characterizing_2019}.

Task models may also take the form of \emph{frameworks}, where the scope, structure, and relationships between of observed tasks, are abstracted into general-purpose hierarchies. Examples include the framework of tasks, sub-tasks, actions, and events proposed by Gotz and Zhou~\cite{gotz_characterizing_2009}, and the goals to tasks framework proposed by Lam et al.~\cite{lam_bridging_2018}.
We observe that these models predict the scope of insights by culling the set of relevant data facts (taxonomies) or narrowing the range of data for applying these data facts (frameworks).
However, these models only represent a range of possibilities. They are inappropriate for describing the \emph{exact} insights an analyst uncovers while completing a visual analysis task.
We take these strengths and limitations into account by defining both a data scope and method scope within our formalism for insights in \autoref{sec:model-design}.

An analyst's interest in pursuing certain tasks can also be defined with respect to the kinds of insights they \emph{expect} to uncover.
This observation stems from the idea that a user's data analysis strategy is likely informed by an initial goal or ``hunch'' regarding the target dataset~\cite{lam_bridging_2018,zgraggen_investigating_2018,battle_characterizing_2019}, even if only vaguely at first~\cite{battle_characterizing_2019}.
For example, Bertin defines tasks according to the structure of the underlying data and the information the user seeks to learn from this data~\cite{bertin1983semiology}. Andrienko and Andrienko extend Bertin's ideas to define tasks as declarative functions over data relations comprised of \emph{targets}, i.e., data attributes of interest, and \emph{constraints}, i.e., query predicates over these attributes~\cite{andrienko2006exploratory}. We note that Andrienko and Andrienko and Bertin's proposals overlap significantly with established definitions of task in database research, notably \emph{relational calculus}, a core component of the \emph{relational model} that also defines tasks (or queries) as declarative functions over data relations~\cite{codd1970relational}. That being said, existing declarative definitions of task are limited to scoping the user's \emph{expectations} and fail to encapsulate the insights that the user actually found, which are often the focus of insight discovery work.

To summarize, existing task models are useful aids for inferring insight scope but are insufficient for fully defining insights. Thus, we focus our formalism on scoping insights directly rather than indirectly through task-based models.
\section{\revised{Formalism Design} Goals}
\label{sec:research-goals}

This paper aims to synthesize existing theories into a unified formalism that represents the core building blocks of insight. Such a formalism could provide theoretical consistency and structure to future insight-based theories and evaluations. To guide \revised{our development of a new formalism}, we summarize three recurring principles from the current theory that clarify the scope, structure, and complexity of insights. Alongside each principle, we propose a research goal that informs our formalism in \autoref{sec:model-design}.

\vspace{3mm}
\noindent
\textit{Principle 1: Insights Represent Linked Units of Knowledge.} We observe in the literature that insights establish \emph{links} between the user's data manipulations, data observations, and their knowledge of related phenomena, and new insights often link back to old ones as analysts' understanding of the data evolves  (\autoref{sec:background:insight:definitions}). As a result, insights are generally complex objects with multiple components. An insight is not just a piece of data but the user’s \emph{interpretation} of this data, which may involve connections with prior knowledge. These components appear consistent across the literature but may vary \revised{in} name/terminology (\autoref{sec:background:insight:categories}).

It is challenging to capture these complex relationships within an atomic insight definition. Instead, we formalize the constituent parts of insight and their relationships.
This idea resembles a well-known problem in visualization specification. Overlaps in visualization taxonomies suggest core components that can be formalized into guiding theory, e.g., the Grammar of Graphics~\cite{wilkinson2012grammar}, which in turn guide the development of visualization grammars such as Vega-Lite~\cite{satyanarayan2017vegalite}.
To this end, we establish our first research goal:

\begin{quote}
    \textbf{Research Goal 1 (RG1):} Rather than defining insights as atomic units, focus on the low-level components of insight
    and their relationships.
\end{quote}

\vspace{3mm}
\noindent
\textit{Principle 2: Insights Distinguish Domain Knowledge From Analytic Knowledge.} The literature suggests that insight is not just the simple linking of two nodes in a graph. Insights occur at specific points within the analysis process, particularly when the user's prior experience intersects with the data at hand in a meaningful way. For this to occur during visual analysis, the visualizations must extend the user's knowledge base to bring in new information (\autoref{sec:background:insight:categories}), such as by revealing new observations or facts about a dataset \revised{(\autoref{sec:background:insight:definitions}).
We define \emph{analytic knowledge} as statistical, visual, or factual information that can be derived directly from the target dataset.  For example, we can observe whether crime is trending up or down in a given city by tallying reported crimes.}
Then, the user can connect these findings with their own \revised{knowledge and experiences} to clarify the significance of \revised{analytical results.
Inspired by the concept of domain knowledge in knowledge modeling research~\cite{yun2021knowledge}, we define \emph{domain knowledge} as contextual information that cannot be inferred from the target dataset. For example, crime datasets may tell us \emph{what} crimes were reported but not \emph{why} they occurred; this additional context requires a deductive approach to analyzing the crime data.}
Insights can also build upon each other, where higher-level insights likely draw on existing links between analytic \revised{(inductive, bottom-up)} and domain \revised{(deductive, top-down)} knowledge (\autoref{sec:background:insight:definitions}). Based on these ideas, we propose our second research goal:

\begin{quote}

\textbf{Research Goal 2 (RG2):} Insights link analytic and domain knowledge, requiring the ability to distinguish between \revised{them while preserving the relationships and patterns that bind them together.}
    
\end{quote}

\vspace{3mm}
\noindent
\textit{Principle 3: Insight Discovery is About the \emph{Interpretation} of Visualizations, Not the Visualizations Themselves.} We observe that studying insight discovery requires more than just collecting the visualizations that users create. If visualization is about insight, not pictures~\cite{hullman2019purpose}, then insight discovery is about tracking how people \emph{interpret} the pictures~\cite{kale2022causal,monadjemi_competing_2020,karer2021insight}, not tracking the pictures themselves nor the corresponding interactions. In other words, recording visualization provenance, i.e., how visualizations are created over time, is a poor substitute for recording actual insights. To understand the relationship between visualizations and insights, we need to be able to track them in equally fine detail. We observe established methods~\cite{xu_survey_2020} and formalisms~\cite{gathani2022grammar} for visualization provenance but a dearth of counterparts for insights, leading to our final research goal:

\begin{quote}
    \textbf{Research Goal 3 (RG3):} Insights represent the \emph{interpretation} of data, e.g., \revised{knowledge} gleaned from a visualization, not just the visualization itself.
\end{quote}


\section{A Formalism for Visualization Insights}
\label{sec:model-design}
\vspace{-1mm}



As a first step towards clarifying the value of existing insight theory, we formalize the theoretical building blocks of insight as agreed upon in the literature. In this section, we define the formal structure of each building block and identify the theories from which they were sourced.

\vspace{-1mm}
\subsection{Initial Building Blocks}
\vspace{-1mm}

Principle 1 highlights the prevailing assumption that insights connect \revised{knowledge}. 
This assumption implies a notion of \emph{links} between knowledge units. Although links are discussed frequently in the visualization literature (see \autoref{sec:background:insight:definitions}), they lack a formalism defining their structure. The closest \revised{we} observe is proposed by Kandogan et al.~\cite{kandogan2018towards}, in which they suggest that knowledge links could be defined using data mining structures, specifically, \emph{knowledge graphs}. Inspired by their approach, we contribute a knowledge graph-driven formalism of links, supporting \textbf{RG1}.
\revised{We draw on existing research in knowledge graph construction from the data mining~\cite{hogan2021knowledge} and knowledge modeling communities~\cite{yun2021knowledge}.}
We \revised{briefly introduce} knowledge graphs from the data mining \revised{and knowledge modeling} literature \revised{and defer to surveys for more details~\cite{hogan2021knowledge,ji2021survey,yun2021knowledge}.}

\vspace{2mm}
\subsubsection{Defining Knowledge Graphs}

A knowledge graph is ``a graph of data intended to accumulate and convey knowledge of the real world, whose nodes represent entities of interest and whose edges represent potentially different relations between these entities''~\cite{hogan2021knowledge}\footnote{Note \revised{that} ``knowledge graph'' and ``knowledge base'' can be used interchangeably \revised{here}. We defer to this survey \cite{ji2021survey} for more details.}.
Entities, the nodes of a knowledge graph, can take various forms, including physical objects such as cities and people~\cite{hogan2021knowledge}, as well as digital objects such as web pages and structured datasets~\cite{bizer2009linked}.
The edges of a knowledge graph are used to record relationships between entities~\cite{hogan2021knowledge,ji2021survey,yun2021knowledge}, such as which cities are located within certain countries or hyperlinks between web pages (e.g., Wikipedia pages) and their backing data (e.g., Wikidata). \revised{In knowledge modeling research, knowledge nodes can also be designed around a framework, where nodes of the same type share the same descriptive attributes defined by the framework~\cite{yun2021knowledge}.}

Visualization research in insight discovery frequently discusses how insights drawn from visualizations can extend a user's knowledge base (see \autoref{sec:background:insight:definitions}), which can be represented as establishing new \emph{nodes} $N$ and \emph{links} (or edges $E$) within a knowledge graph $G(N,E)$.
To capture this agreement, we treat analytic and domain knowledge as \emph{knowledge nodes} $n \in N$ within our formalism. Knowledge nodes can connect to other nodes of the same type, for example, by specifying edges $e \in E$ between related domain knowledge nodes. We also define higher-level nodes that group domain and analytic knowledge nodes together, such as to specify a higher level \emph{insight node}. Together, \emph{knowledge nodes} and \emph{knowledge links} represent our initial building blocks for defining insights. 

\vspace{2mm}
\subsubsection{Defining Knowledge Nodes and Links}
\label{sec:model-design:knowledge-nodes}

\begin{figure}
    \centering
    \includegraphics[width=1.0\columnwidth]{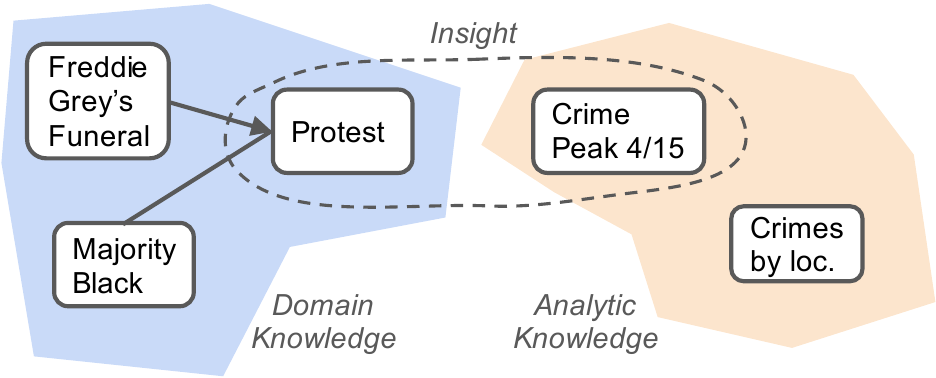}
    \vspace{-2mm}
    \caption{Knowledge nodes can be linked to form insights, e.g., linking observations of peaks in total crimes with related characteristics and historical events for Baltimore. Undirected edges capture relatedness. \revised{Directed} edges record trajectories of knowledge expansion \revised{in} the graph.}
    \label{fig:graph-example}
\end{figure}

\begin{figure}
    \centering
    \includegraphics[width=0.7\columnwidth]{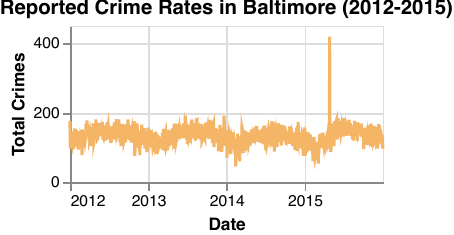}
    \vspace{-2mm}
    \caption{\revised{A peak in reported crimes is observed in April 2015.}}
    \label{fig:crime-peak}
\end{figure}

Our formalism uses a graph-based data model to represent insights, where the nodes $N$ in the graph represent units of (explicit~\cite{nonaka2007knowledge}) knowledge, and the edges $E$ represent the relationship between nodes, which may be directed or undirected.
An example is shown in \autoref{fig:graph-example}, which is based on the Baltimore crime data exploration example by Mathisen et al.~\cite{mathisen_insideinsights_2019}.
We see that a knowledge node can record historical or domain-specific knowledge about related concepts, such as \node{majority black} (``The majority of people living in Baltimore, MD are Black or African American'').
Similarly, we can record knowledge gained by analyzing Baltimore crime data \revised{as shown in \autoref{fig:crime-peak}} (\eg\ \node{Crime Peak 04/15} ``There was a spike in reported crimes in April 2015'').
The edges in the graph represent relatedness and can track the evolution of knowledge acquisition. For example, an undirected edge can indicate the relatedness of \node{Protest} --- \node{Crime Peak 04/15} or a directed edge can indicate that the protests were in response to the unwarranted killing of Freddy Grey, an unarmed black man, by the Baltimore Police  \node{Freddy Grey's Funeral} ---$>$ \node{Protest}. Similarly, if analyzing total crimes by location (on the street, in the home, etc.) turns out to be a ``dead end'' then this analysis may not extend the user's knowledge base, resulting in a node \node{Crimes by loc.} with no edges in the graph. Furthermore, rather than linking protests directly with crime peaks in the graph, we could group them into a higher-level graph node, such as an insight node, shown as a dashed line in \autoref{fig:graph-example}.

We define knowledge links as pairs of knowledge nodes:

\begin{equation*}
    \begin{aligned}
        \mathtt{KnowledgeLink} :=\ & (n_i,n_j),\ n_i \in N, n_j \in N
    \end{aligned}
\end{equation*}
For directed links, we can further specify \texttt{source} and \texttt{target} nodes as part of the knowledge link:
\begin{equation*}
    \begin{aligned}
        \mathtt{DirectedKnowledgeLink} :=\ & (\mathtt{source},\mathtt{target}) \\
        \mathtt{source} :=\ & n_i \in N \\
        \mathtt{target} :=\ & n_j \in N \\
    \end{aligned}
\end{equation*}
To streamline our formalism, we combine knowledge nodes and links within a single definition for \emph{knowledge nodes}:

\begin{equation}
\label{eq:knowledge-node}
\begin{aligned}
 \mathtt{KnowledgeNode} :=\ & \{name,\ \mathtt{sources},\ \mathtt{targets},\ \mathtt{related} \} \\
\mathtt{sources} :=\ & \{ n_i,\ n_j,\ ... \} \subseteq N \\
\mathtt{targets} :=\ & \{n_k,\ n_l,\ ...\} \subseteq N \\
\mathtt{related} :=\ & \{n_p,\ n_q,\ ...\} \subseteq N
\end{aligned}
\end{equation}

where \emph{name} is a string identifier for the specified knowledge node, and other nodes that contributed directly to the current node can be specified as a set of \texttt{sources}. Nodes influenced by the current node can be specified as a set of \texttt{targets}. We can specify nodes in an undirected relationship as a set of \texttt{related} nodes. In the remainder of the paper, we refer only to our definition of knowledge nodes, which implicitly contains our definition of knowledge links.

\vspace{-1mm}
\subsection{Formalizing Domain Knowledge}
\label{sec:model-design:domain-knowledge}
\vspace{-1mm}

Revisiting Principle 2, we observe an emphasis in the literature on distinguishing different types of knowledge, such as \revised{meaningful statistical or mathematical features that can be derived from a target dataset, i.e., \emph{analytic knowledge}, and contextual information (e.g., personal experience, domain expertise) that cannot be derived from this dataset, i.e., \emph{domain knowledge}.}
\revised{In} \autoref{fig:graph-example}, events such as Freddy Gray's funeral represent domain knowledge regarding the history of Baltimore (highlighted in blue). The statistics calculated using Baltimore crime reports represent analytic knowledge (highlighted in orange), e.g., that April 2015 contained some of the highest days of reported crimes between 2012 and 2015.
In our formalism, we capture this distinction by formalizing the defining characteristics of domain knowledge and analytic knowledge nodes, supporting \textbf{RG2}.

We integrate existing ideas to formalize our collective understanding of domain knowledge.
Gotz et al.~\cite{gotz_interactive_2006} suggest that analysts reason about different \textit{concepts} as they analyze a dataset, as well as specific \textit{instances} of these concepts that stand out.
Extending these ideas, we treat \emph{concepts} as custom type definitions representing a belief or idea that the analyst wants to express, such as the concepts of ``racism'', ``conspiracy'', or ``protest''.
We define an \texttt{instance} as a representative case
of a \emph{concept} and the supporting data \texttt{T}:

\begin{equation}
\label{eq:instance}
\begin{aligned}
\mathtt{instance} :=\ & \{name,\ concept,\ \mathtt{T}\} \\
\mathtt{T} :=\ & \{a_1,\ a_2,\ ...\}
\end{aligned}
\end{equation}
where \texttt{T} is a relational table with at least one data attribute/column ($a_1$, $a_2$, etc.) and row. Revisiting our running example, suppose we learn about the 2015 Baltimore Protests through Wikipedia and Vox. We can record this in \texttt{T} using two attributes ($a_1=$ \texttt{Source}, $a_2=$ \texttt{Link}) and two rows for the corresponding articles, shown in \autoref{tab:instance-t}.

\begin{table}
\centering
\caption{Example table \texttt{T} for creating an \texttt{instance} representing the 2015 Baltimore Protests using \autoref{eq:instance}.}
\label{tab:instance-t}
 { \scriptsize
\begin{tabular}{ll}
\hline
\textbf{Source} & \textbf{Link}                               \\ \hline
Wikipedia       & \href{https://en.wikipedia.org/wiki/2015_Baltimore_protests}{\texttt{https://en.wikipedia.org/wiki/2015\_Baltimo...}} \\
Vox             & \href{https://www.vox.com/2016/7/27/18089352/freddie-gray-baltimore-riots-police-violence}{\texttt{https://www.vox.com/2016/7/27/18089352/fre...}}  \\ \hline
\end{tabular}
}
\end{table}

\revised{Using the concept of \textit{frames} from the knowledge modeling literature~\cite{yun2021knowledge}, we} formally define a \texttt{DomainKnowledgeNode} \revised{by extending} our \texttt{KnowledgeNode} definition from \autoref{eq:knowledge-node} to include properties from our \texttt{instance} definition in \autoref{eq:instance}:
\begin{equation}
\label{eq:domain-knowledge-node}
\begin{aligned}
 \mathtt{DomainKnowledgeNode} :=\ & \{name,\ \mathtt{sources},\ \mathtt{targets}, \\
     & \ \mathtt{related},\ concept,\ \mathtt{T} \}
\end{aligned}
\end{equation}

In this way, we can capture meaningful concepts and instances expressed within a unit of domain knowledge as well as track their propagation through a knowledge graph.

\vspace{-1mm}
\subsection{Formalizing Analytic Knowledge}
\label{sec:model-design:analytic-knowledge}
\vspace{-1mm}

Analytic knowledge lacks a precise definition in the literature. For example,
Gotz et al. appear to define analytic knowledge as annotations to domain knowledge~\cite{gotz_interactive_2006}. 
Alternative definitions proposed by Shrinivasan and van Wijk~\cite{shrinivasan_supporting_2008} and Andrienko and Andrienko~\cite{andrienko_viewing_2018} seem to suggest that analytic knowledge is information gained from querying and interacting with a dataset. These ideas also inspired our third research goal (\textbf{RG3}): to distinguish users' creation of visualizations from their interpretation of visualizations.
\revised{In \autoref{sec:research-goals}, we define
analytic knowledge as information derived through the manipulation of data} such as through interacting with visualizations.
Towards formalizing this \revised{definition}, we first clarify what we mean by ``information derived'' (i.e., \emph{data relationships}) from ``manipulation of data'' (i.e., \emph{data transformations}).

\vspace{2mm}
\subsubsection{Data Relationships}
\label{sec:model-design:data-relationship}

We deduce from the prior work that when researchers encounter analytic knowledge in insight-based studies (see \autoref{sec:background:insight:definitions}), they tend to interpret it in terms of \emph{mathematical or statistical characteristics}, such as by recording associated data correlations, distributions, patterns, or anomalies observed by users~\cite{saraiya_insight-based_2005,yi_understanding_2008,zgraggen2017progressive,yang_chen_toward_2009,choe_characterizing_2015}.
We refer to these characteristics as \revised{\textbf{\emph{data relationships}}}, but they are also referred to as data facts in the literature (see \autoref{sec:background:insight:definitions}). One could simply try to track the visualizations a user created as a proxy for data relationships. However, users can easily draw different conclusions from the same visualizations~\cite{zehrungsinghal2021vis}, making visualizations ambiguous records of analytic knowledge~\cite{kale2022causal}. Our goal is not to collect pictures but to understand the knowledge gleaned from them. For these reasons, we define analytic knowledge in terms of \emph{properties that can be calculated or modeled directly from the underlying dataset}. We stress that our formalism represents a \emph{quantitative interpretation} of the analytic knowledge gained and not the raw utterances of study participants.

There are many ways in which researchers have interpreted data relationships, such as by creating statistical representations like histograms~\cite{battle2016dynamic}, linear regression models~\cite{harrison2014ranking} or statistical sketches~\cite{demiralp_foresight_2017,yang_chen_toward_2009} or even machine learning models like Hidden Markov models~\cite{monadjemi_competing_2020,ottley_follow_2019} or support vector machines~\cite{battle2016dynamic}. Amid this panoply of techniques, we observe a recurring high-level structure that we leverage to formally define the theoretical properties of data relationships.

Given a relational table \texttt{T} with data attributes $\mathtt{A} = \{ a_1, a_2, ..., a_n \}$, we find that \textbf{\emph{multivariate data relationships}}: (1) take one or more data attributes as input variables for training and one attribute as an output variable for prediction; (2) define a training function to build a model that predicts the output given the input; and (3) define a prediction function that uses the model to map new inputs to projected outputs:
\begin{equation}
\label{eq:data-relationship-multi}
\begin{aligned}
 \mathtt{MultivariateRelationship} :=\ & \{ f_{train},\ f_{predict} \} \\
 f_{train}/f_{predict} :=\ & \mathtt{A}_{ij} \mapsto a_o,\ \mathtt{A}_{ij} \subset \mathtt{A},\\
  & a_o \in \mathtt{A}
\end{aligned}
\end{equation}
The training function can be as simple as calculating coefficients for a linear equation, but it can also be complex, such as training machine learning models. We find that \textbf{\emph{univariate relationships}} such as kernel density estimation: (1) take a single input attribute to simulate; (2) define a function to train a model to capture the corresponding distribution; and (3) define a function to simulate records from the modeled distribution:
\revised{
\begin{equation}
\label{eq:data-relationship-uni}
\begin{aligned}
 \mathtt{UnivariateRelationship} :=\ & \{ f_{train}, f_{simulate} \} \\
 f_{train} :=\ & a_i \mapsto \emptyset, a_i \in \mathtt{A}\\
 f_{simulate} :=\ & \emptyset \mapsto a_i
\end{aligned}
\end{equation}
}

This definition covers all of the data relationships observed in our literature review, and easily extends to new data relationships.
For example, any new multivariate relationship that takes one or more attributes as input, predicts an attribute as output and provides the requisite function types are automatically covered under our formalism.

\vspace{2mm}
\subsubsection{Data Transformations}
\label{sec:model-design:data-transformations}

Typically, analysts \revised{must} process their data to \revised{facilitate} insight discovery~\cite{stolte2002polaris,shneiderman_eyes_1996}, such as by filtering, aggregating, sorting, etc., which we refer to as \textbf{\emph{data transformations}}.
For example, to answer the question \emph{``which $k$ dates have the most reported crimes?''} we have to group the Baltimore crime data by date, count all reported crimes per date, then sort by the count to retrieve the top $k$ dates.

Data transformations can have profound effects on an analyst's ability to extract insights. For example, Battle and Heer found that differences in users' interaction sequences in Tableau could lead to different queries being executed \revised{over the} data and ultimately different answers to the same analysis tasks~\cite{battle_characterizing_2019}.
Furthermore, interfaces that hinder interactive data processing have been shown to negatively impact insight \revised{generation}~\cite{liu_effects_2014,zgraggen2017progressive}.
Given the critical role \revised{of data transformations in insight discovery, we consider them essential to formalizing analytic knowledge}.

Similar to Andrienko and Andrienko~\cite{andrienko2006exploratory}, we specify data transformations as queries over relational tables. We use \emph{relational algebra} to represent data transformations, where relational algebra can be considered the dual to \emph{relational calculus}~\cite{codd1970relational} (see \autoref{sec:background:insight:scope} for more on relational calculus). To do this, we treat data transformations as a sequence of relational algebra operations, where each operation is essentially a function that takes a relational table \texttt{T} as input and returns a relational table $\mathtt{T}'$ as output:
\begin{equation}
\label{eq:data-transformation}
\begin{aligned}
 \mathtt{DataTransformation} :=\ & [o_1, o_2, ..., o_i, ...] \\
 o_i :=\ & \mathtt{T} \mapsto \mathtt{T}'
\end{aligned}
\end{equation}
where $\mathtt{T}$ may not have the same attributes \revised{or rows as $\mathtt{T}'$. For example, to calculate \revised{peak crime dates in Baltimore}, we: group by date, count reported crimes per date, and sort the dates by count. The aggregation produces fewer attributes and rows than the original input table since we are grouping all reported crimes by date and returning a single count per date. In contrast, the sort returns an output table with the same shape as its input}, since the rows are simply reordered.

As hinted at by Bertin~\cite{bertin1983semiology} and Andrienko and Andrienko~\cite{andrienko2006exploratory}, tracking data transformations can provide an advantage over interaction logs, since it emphasizes a user's interpretation of the data rather than the idiosyncrasies of a particular user interface. For example, Tableau desktop offers a myriad of ways to filter a dataset through its interface but they all map to the same filtering operations in relational algebra ~\cite{stolte2002polaris,battle_characterizing_2019}.
\revised{Although} our formalism differs from traditional SQL, it matches related languages such as Microsoft LINQ~\cite{meijer2011world}, which can be mapped to SQL~\cite{chandramouli2014trill}.

\vspace{2mm}
\subsubsection{Analytic Knowledge Nodes}

Similar to \revised{domain knowledge nodes (\autoref{sec:model-design:domain-knowledge})}, we formalize analytic knowledge nodes by merging our base definition for knowledge nodes (\autoref{eq:knowledge-node}) with our definitions for data transformations (\autoref{eq:data-transformation}) and data relationships \revised{(\autoref{eq:data-relationship-multi} and \autoref{eq:data-relationship-uni})}:
\begin{equation}
\label{eq:analytic-knowledge-node}
\begin{aligned}
 \mathtt{AnalyticKnowledgeNode} :=\ & \{name,\ \mathtt{sources},\ \mathtt{targets}, \\
     & \ \mathtt{related}, \\
     & \ \mathtt{dataTransformation}, \\
     & \ \mathtt{dataRelationship} \} \\
\mathtt{dataTransformation} :=\ & [o_1, o_2, ..., o_i, ...] \\
\mathtt{dataRelationship} :=\ & \{ f_{train},\ f_{predict} \} \\
     & |\ \{ f_{train},\ f_{simulate} \} \\
\end{aligned}
\end{equation}

With this definition, researchers gain access to a precise, quantitative representation of a user's analytic knowledge. We record exactly how the data has been manipulated and can quantify the knowledge we believe the user gained from their results. Further, this representation remains consistent regardless of differences in user interfaces, enabling it to generalize across visualization tools.

\vspace{-1mm}
\subsection{Formalizing Insights}
\label{sec:model-design:insight}
\vspace{-1mm}

\begin{figure}[h]
    \centering
    \includegraphics[width=0.9\columnwidth]{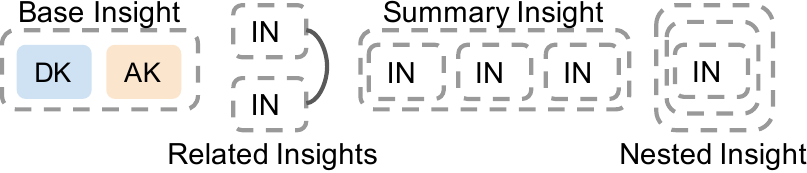}
    \vspace{-3mm}
    \caption{\revised{Various insight structures enabled by our formalism. ``DK'' is domain knowledge, ``AK'' analytic knowledge, and ``IN'' insight.}}
    \label{fig:insight-structures}
\end{figure}

With \revised{precise definitions for domain (\autoref{eq:domain-knowledge-node}) and analytic (\autoref{eq:analytic-knowledge-node}) knowledge, we complete our formalism for insights.} As shown in \autoref{fig:graph-example}, we view insights as a \emph{cluster} of relevant domain and analytic knowledge nodes within the user's knowledge graph. However, simply drawing a link between the target nodes renders this \emph{hierarchical} relationship indistinguishable from other relationships in the graph. To express the hierarchical nature of insight, we define it as a \emph{higher level representation} of knowledge:
\begin{equation}
\label{eq:insight-node}
\begin{aligned}
 \mathtt{InsightNode} :=\ & \{name,\ \mathtt{sources},\ \mathtt{targets}, \\
     & \ \mathtt{related},\ \mathtt{domainKnowledge}, \\
     & \ \mathtt{analyticKnowledge} \} \\
\mathtt{domainKnowledge} :=\ & \{ \mathtt{domainKnowledgeNode_1}, \\
    & \ \mathtt{domainKnowledgeNode_2,...} \} \\
\mathtt{analyticKnowledge} :=\ & \{ \mathtt{analyticKnowledgeNode_1}, \\
    & \ \mathtt{analyticKnowledgeNode_2},... \} \\
\end{aligned}
\end{equation}
where more than one node can be specified in \revised{\texttt{domainKnowledge} and \texttt{analyticKnowledge} to support more complex insights. Insights can also be hierarchical by integrating other insights~\cite{pirolli_sensemaking_2005,smuc_score_2009,gotz_interactive_2006} (see \autoref{fig:insight-structures}).}

\revised{To support these relationships, we define insights as higher level \emph{graph nodes} that inherit from \texttt{KnowledgeNode} (\autoref{eq:knowledge-node}), supporting linking and extension of insights as the user's knowledge graph evolves over time (see \autoref{fig:insight-structures}). 
Using our formalism, we can link any previous insights as \texttt{sources} that informed the current insight and new insights as \texttt{targets} that were informed by the current insight.}


\vspace{-1mm}
\subsection{Implementation}
\vspace{-1mm}

To support transparency, evaluation, and reuse of our formalism by the research community, we have implemented a specification language called \textsc{\model} that is freely available online. \model\ can be used to specify \emph{all} of the building blocks defined above: concepts, instances, domain knowledge, data transformations, data relationships, analytic knowledge, and insights. \model\ is implemented in TypeScript and can be used in TypeScript and JavaScript projects. \model\ supports both the Node and Observable JavaScript environments.

Rather than re-implementing \revised{known data transformations, we import existing libraries into \model\ using wrapper classes matching} the specifications in \autoref{sec:model-design:data-relationship} and \autoref{sec:model-design:data-transformations}.
Our codebase imports the Vega \revised{transforms~\cite{vega2023transforms} and Arquero~\cite{heer2021arquero} libraries}, so \revised{all data transformations supported by Vega or Arquero can also be used} in \model. Likewise, we import a wide range of existing data relationships, such as linear regression models,
anomaly detection through isolation forests,
and univariate distributions via kernel density estimation through wrapper classes that integrate existing libraries.


\begin{figure}[b]
    \centering
    \includegraphics[width=\columnwidth]{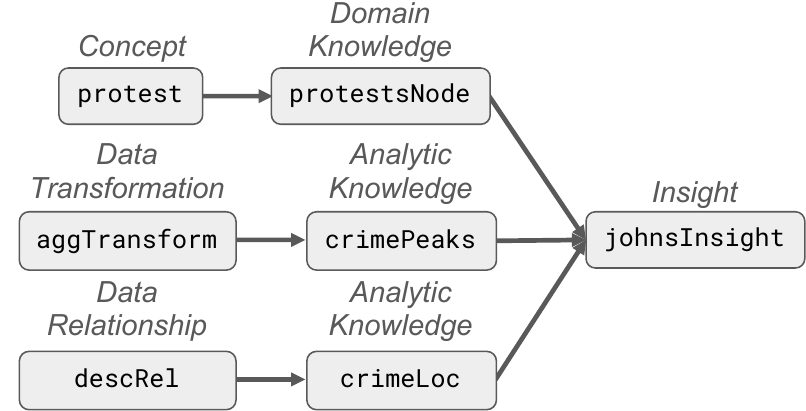}
    \caption{\revised{An implementation of our running example from \autoref{sec:model-design}. Each node represents a specified \model\ object and its corresponding component from our formalism. The directed edges represent input relationships. For example, the \texttt{protest} object is an input to the \texttt{protestsNode} object.}}
    \label{fig:crimetask-overview}
\end{figure}

\section{\revised{Practical Use Cases}}
\label{sec:case-studies}
\vspace{-1mm}


\revised{In this section, we present \revised{four use cases} demonstrating how to apply \model.}
We also highlight future research opportunities enabled by our formalism.
The corresponding \model\ code is shared in our supplemental materials.


\subsection{\revised{Use Case 1: Recreating an Analysis Session}}

\revised{In this use case, we show how to use \model\ by recreating our ongoing example from Mathisen et al.~\cite{mathisen_insideinsights_2019}, which follows a fictional analyst, John, as he investigates crime peaks in Baltimore from 2012 through 2015. John's analysis uncovers the first crime peak on April 27, 2015 which coincides with protests sparked by the funeral of Freddie Gray, a young black man who was killed by the Baltimore police. An overview is provided in \autoref{fig:crimetask-overview}, where each node represents a particular \model{} object that we create throughout the example, and the edges between nodes represent relationships between objects.}

\subsubsection{\revised{Specifying Domain Knowledge}}

\revised{\model\ enables us to formalize John's domain knowledge as abstract \emph{concepts}, e.g., the ``protests'' and specific \emph{instances} of these concepts, e.g., the ``Baltimore protests.'' We define the concept ``Protest'' (lines 1-4). Then, we define an instance of the ``Protest'' concept based on the Baltimore Protests in 2015 using a new domain knowledge node (lines 5-21).}

\vspace{1em}
\begin{minipage}{.9\linewidth}
\begin{minted}
{javascript}
const protest = new Concept(
  "Protest", // name
  [] // parentConcepts
);
const protestsNode = new DomainKnowledgeNode(
    "2015BaltimoreProtests", // name
    protest, // associated concept
    { // metadata
        attributes: [{
            name: "Source",
            type: nominal
        }, {
            name: "Link",
            type: nominal
        }],
        values: [{
            "Source": "Wikipedia"
            "Link": "https://en.wikip..."
        }]
    }
);
\end{minted}
\end{minipage}
\vspace{1em}

\revised{Since domain knowledge inherits from knowledge nodes (see \autoref{sec:model-design:domain-knowledge}), we can also specify node relationships, such as one node ``causing'' or being ``related to'' another node. For example, to link a source (i.e., parent) node, we call the \texttt{addSource} function on the \texttt{protestsNode} object.
}

\subsubsection{\revised{Specifying Analytic Knowledge}}
\revised{To develop analytic knowledge, analysts infer \emph{data relationships} (\autoref{sec:model-design:data-relationship}). To prepare the data for visualization or modeling, analysts apply \emph{data transformations} (\autoref{sec:model-design:data-transformations}). We demonstrate how to create data transformations by using Arquero to calculate total reported crimes per day and identify peak crime days.}


\vspace{1em}
\begin{minipage}{.9\linewidth}
\begin{minted}
{javascript}
// Arquero Data Transformation
const aggTransform = {
    sources: [baltimoreCrime],
    transforms: [
      // group by day
      { op: "groupby", args: ["CrimeDate"] },
      // count crimes per day
      { op: "rollup", args: [{ count: op.count() }] },
      // sort days by count
      { op: "orderby", args: [desc("count")] },
      // return top 3 days with highest counts
      { op: "filter", args: [() => op.rank() <= 2] }
    ]
};
 const getAggTransformResults =
  () => executeDataTransformation(aggTransform);
\end{minted}
\end{minipage}
\vspace{1em}

\revised{First, we specify the input dataset (line 3). Then, we group reported crimes by date (line 6), count total records per day (line 8), sort the dates by count (line 10), and filter for the top three dates with the highest counts (line 12). Finally, we execute the transformations using the \texttt{executeDataTransformation} method on line 16.}


\vspace{1em}
\begin{minipage}{.9\linewidth}
\begin{minted}
{javascript}
const crimePeaks = new AnalyticKnowledgeNode(
  "peakCrimes", // node name
  Date.now(), // timestamp
  aggTransform, // data transformation
  null, // data relationship
  getAggTransformResults, // results
);
\end{minted}
\end{minipage}
\vspace{1em}

\revised{Next, we record how John processed the data to identify peaks in a new analytic knowledge node.
First, we give this analytic knowledge a name (line 2) and record when John learned it (line 3). Then, we connect relevant data transformations and/or relationships (lines 4-5). In this case, John's findings relate only to the \texttt{aggTransform} object. 
Similar to domain knowledge nodes, we can link to other analytic knowledge nodes using \texttt{addSource}, \texttt{addTarget}, and \texttt{addRelated}. All node objects share this property.}

\revised{To demonstrate data relationships in \model\, consider this extension of the Baltimore example: suppose John is curious whether location is indicative of crime type, for example, whether different crimes happen indoors versus outdoors, or in an apartment versus a business. We can specify a new model to predict this relationship as follows:}

\vspace{1em}
\begin{minipage}{.9\linewidth}
\begin{minted}
{javascript}
const descRel = new DecisionTreeClassification(
  "predictCrimeType", // name
  [ // input attributes to predict with
    {
      name: "Inside/Outside",
      attributeType: AttributeType.nominal
    },
    {
      name: "Premise",
      attributeType: AttributeType.nominal
    }
  ],
  // output attribute to be predicted
  {
    name: "Description",
    attributeType: AttributeType.nominal
  }
);
dt.train(baltimoreCrimes.records);
const prediction = dt.predict(baltimoreCrimes.records[0]);
\end{minted}
\end{minipage}
\vspace{1em}

\revised{Since the input and output attributes are categorical, we specify a decision tree classifier to predict their relationship (line 1). The input attributes used to train the model are ``Inside/Outside'' and ``Premise'' (lines 3-12). The output attribute being predicted is ``Description'' (lines 14-17), which describes the type of crime reported. However, using a machine learning relationship is not required, and the model type can easily be swapped in \model{} by choosing a relationship type other than \texttt{DecisionTreeClassification}.
Line 19 shows how we can train the specified decision tree model on the Baltimore crimes dataset and Line 20 shows how this model can be used to predict the crime type of specific records. In this way, specified analytic knowledge can be evaluated for statistical rigor by testing the accuracy of the underlying data relationships, supporting prior calls for more precise evaluation of user insights~\cite{zgraggen_investigating_2018}.}

\revised{Suppose that \texttt{Inside/Outside} and \texttt{Premise} are not strong predictors of crime \texttt{Description}. We can record this result in a new AnalyticKnowledgeNode as follows:}

\vspace{1em}
\begin{minipage}{.9\linewidth}
\begin{minted}
{javascript}
const crimeLoc = new AnalyticKnowledgeNode(
  "crimeLocations", // node name
  Date.now(), // timestamp
  null, // data transformation
  descRel, // data relationship
  null // results
);
\end{minted}
\end{minipage}
\vspace{1em}

\revised{\model\ supports any type of multivariate relationship, including K nearest neighbors, linear regression, and naive Bayes models, as well as univariate relationships (e.g., via kernel density estimation) and other statistical relationships such as outliers (e.g., via isolation forests).}

\subsubsection{\revised{Specifying Insight}}
\revised{The last step in specifying insights is to link domain knowledge with relevant analytic knowledge. Continuing our example, we specify an insight connecting John's domain knowledge about the Baltimore protests (line 4) and analytic knowledge regarding peak crime dates in Baltimore (line 5):}

\vspace{1em}
\begin{minipage}{.9\linewidth}
\begin{minted}
{javascript}
const johnsInsight = new InsightNode(
  "johnsInsight", // name
  // domain knowledge
  [protestsNode], 
  // analytic knowledge
  [crimePeaks, crimeLoc]
);
\end{minted}
\end{minipage}
\vspace{1em}

\revised{We can also keep track of John's ``dead ends'' as desired, for example including our \texttt{crimeLoc} analytic knowledge node as shown on line 4. Similar to the domain and analytic knowledge nodes, insight objects also support the linking of source, target, and related insights.}

\vspace{-1mm}
\subsection{\revised{Use Case 2}: Analyzing Insight Complexity}
\label{sec:case-studies:complexity}
\vspace{-1mm}

\revised{Similar to insight scope, insight complexity is an} important concept in the literature. For example, insight complexity is \revised{used to estimate the quality or value of} insights~\cite{north_toward_2006,smuc_score_2009,he_characterizing_2020}.
In this \revised{use case}, we explore how North defines insight complexity~\cite{north_toward_2006} and connect this definition with alternative conceptualizations of insight complexity in the literature.

\begin{figure*}
    \centering
    \includegraphics[width=1.0\textwidth]{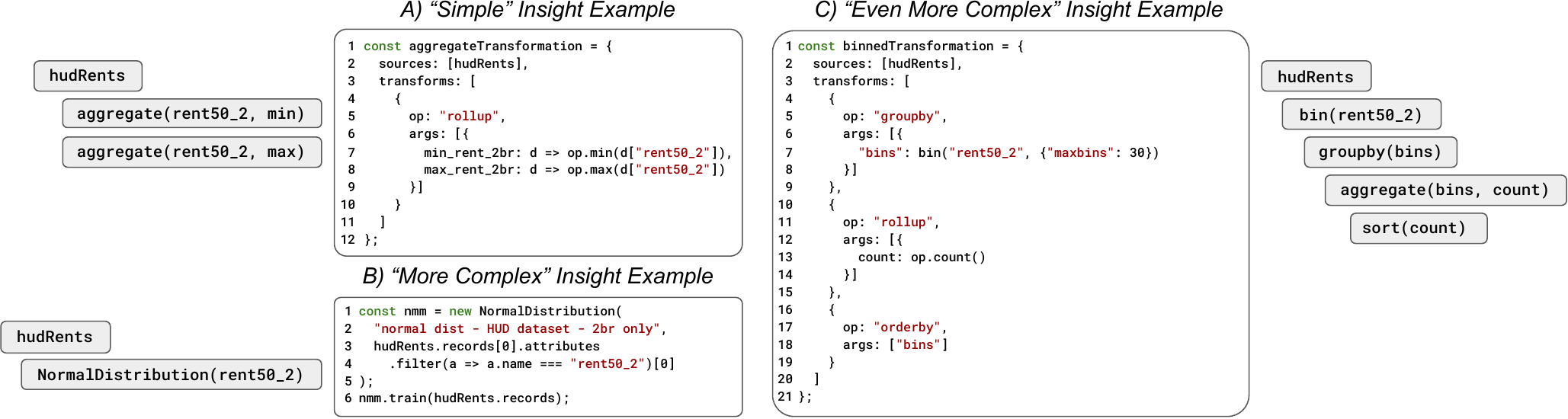}
    \caption{\revised{North's three levels of insight complexity~\cite{north_toward_2006} in \model\ alongside the corresponding relational operators.}}
    \label{fig:north-example}
\end{figure*}

\vspace{2mm}
\subsubsection{Three Levels of Complexity} North argues that ``complexity is determined by how much data is involved in the insight'' \revised{and gives} three forms of ``insights'' over monthly rents \revised{as examples}~\cite{north_toward_2006}.
\revised{We} implement these insights \revised{(see \autoref{fig:north-example}) using data from the U.S. Department of Housing and Urban Development}\footnote{\url{https://www.huduser.gov/portal/datasets/50per.html}}.

\begin{enumerate}[itemsep=0em,topsep=0em, label=\Alph*)]
\item \emph{``Simple'' Insights.} North's simplest insight computes minimum and maximum rent values, \revised{which we calculate} using an aggregate data transformation.

\item \emph{``More Complex'' Insights.} North proposes a ``more complex'' insight that estimates a normal distribution over the rent data, which we calculate as a univariate data relationship using Vega's statistics package\footnote{\url{https://github.com/vega/vega-statistics/}}.

\item \emph{``Even More Complex'' Insights.} The most complex insight estimates the shape of the rent distribution using a histogram. This \revised{maps} to a series of data transformations to bin the data and then aggregate it by bin ranges.
\end{enumerate}

\vspace{2mm}
\subsubsection{Insight Complexity $\subseteq$ Query Complexity?} 
\revised{According} to our formalism, North's examples involve a univariate data relationship (a normal distribution) and two sets of data transformations (min/max calculations and binned aggregation). Further, North seems to emphasize the data transformations, which are essentially relational queries over data (see \autoref{sec:model-design:analytic-knowledge}).
\revised{We see this in \autoref{fig:north-example}, where binned aggregation requires a more complex sequence of operations compared to the min/max calculations.}
\revised{Here,} ``insight complexity'' could mean \emph{relational query complexity}, which aligns well with existing definitions of insight posed by Andrienko and Andrienko~\cite{andrienko2006exploratory} and Bertin~\cite{bertin1983semiology} as well as relevant database research~\cite{jain2016sqlshare} and even research on program complexity~\cite{mccabe1976complexity}. In this sense, the Andrienko and Andrienko~\cite{andrienko2006exploratory} and Bertin~\cite{bertin1983semiology} definitions could be considered generalizations of North's definition of insight complexity since relational queries can encompass a wider range of programs than min-max or histogram calculations. Queries also provide an easy means of calculating data coverage, which North uses to define complexity.

That being said, defining insight complexity as query complexity makes data manipulation the focus of insight rather than how people interpret data, i.e., as \emph{data relationships}. Thus, this view of insight complexity is deficient, as hinted at by North's inclusion of a univariate data relationship. Saraiya et al.~\cite{saraiya_insight-based_2005} and Smuc et al.~\cite{smuc_score_2009} extend this idea to include multivariate relationships in their definitions. Kandogan and Engelke take this a step further by applying relational query patterns to express data relationships such as linear correlations~\cite{kandogan2018towards}, further enriching our understanding of insight complexity. With a richer definition of insights comes the need for alternative methods for measuring insight complexity, which could be an exciting opportunity for future research.

\vspace{2mm}
\subsubsection{\revised{Benefits of the Formalism}} Using our formalism, we could hypothesize new measures of insight complexity, such as by measuring knowledge \emph{depth} by computing the longest path from a knowledge node to its earliest ancestor, which aligns with existing definitions of exploration depth~\cite{battle_characterizing_2019}. Using \model{}, knowledge node depth can be calculated programmatically by backtracking from a node's directed edges. This calculation can also be augmented to include the number of relational operations involved in connected data transformations.
Consistent with North's assertion, we could even measure knowledge \emph{breadth} as the percentage of data values (rows $\times$ attributes/columns) involved as inputs and outputs to the corresponding data transformations and/or relationships.

\begin{figure*}
    \centering
       \centering
    \begin{tabular}[t]{cc}
\begin{subfigure}{0.6\textwidth}
    \centering
    \smallskip
    \includegraphics[width=\textwidth]{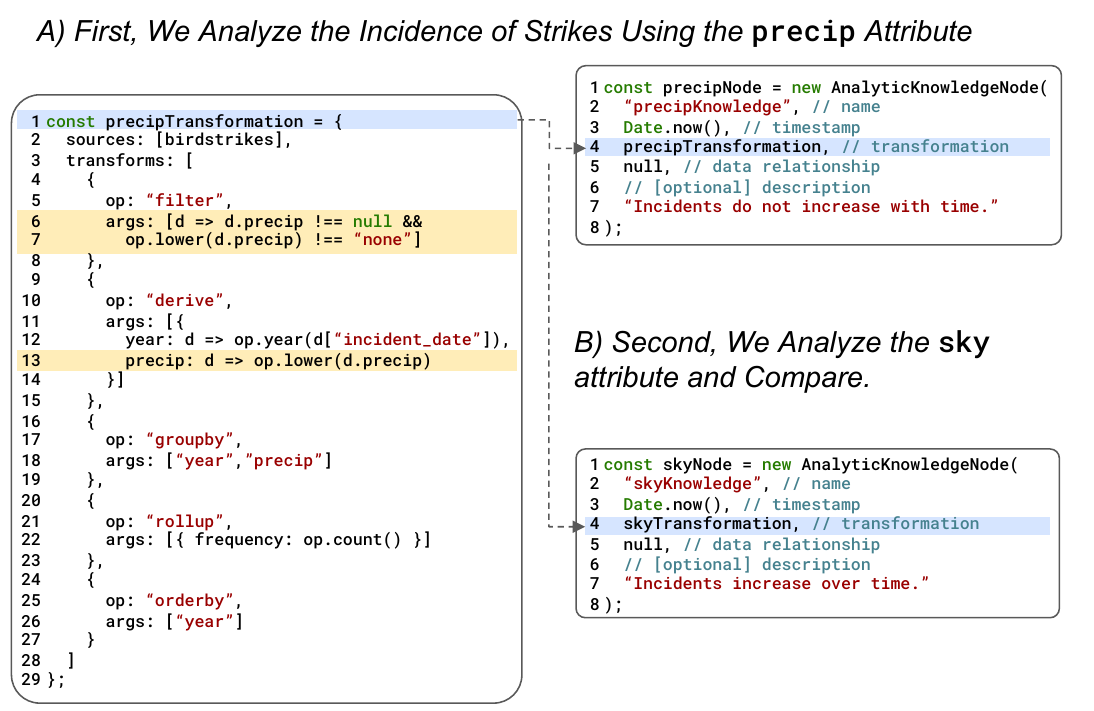}
    \caption{\revised{Specified data transformations (left) and analytic knowledge nodes (right). Only lines 6-7 and 13 need to change to analyze the \texttt{precip} or \texttt{sky} attribute, shown in yellow.}}
    \label{fig:case-studies:birdstrikes-code}
\end{subfigure}
    &
        \begin{tabular}{c}
        \smallskip
            \begin{subfigure}[t]{0.36\textwidth}
                \centering
                \vspace{-80mm}
                \includegraphics[width=\textwidth]{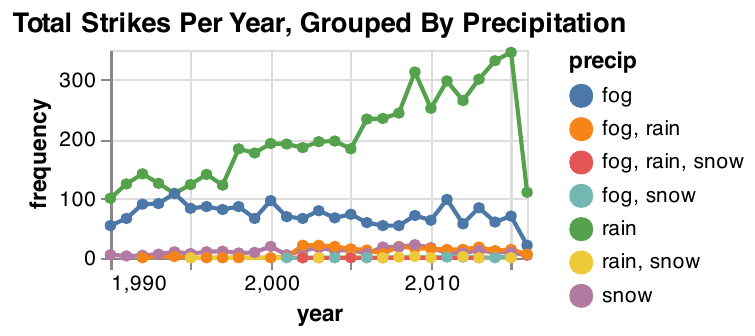}
                \vspace{-5mm}
                \caption{\revised{Analyzing the \texttt{precip} attribute. Strikes \emph{do not} appear to increase with time.}\vspace{-10mm}}
                \label{fig:case-studies:precip}
            \end{subfigure}\\
            \begin{subfigure}[t]{0.36\textwidth}
                \centering
                \vspace{-40mm}
                \includegraphics[width=\textwidth]{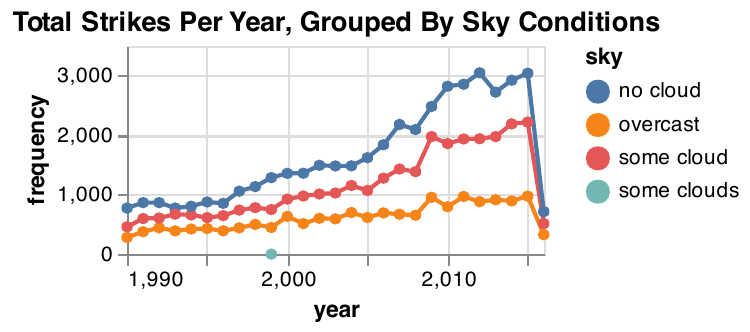}
                \vspace{-5mm}
                \caption{\revised{Analyzing  the \texttt{sky} attribute. Strikes appear to \emph{increase} with time.}}
                \label{fig:case-studies:sky}
            \end{subfigure}
        \end{tabular}\\
    \end{tabular}
    \vspace{-4mm}
    \caption{\revised{In a prior study~\cite{battle_characterizing_2019}, participants explored wildlife-aircraft strikes under various weather conditions, \texttt{precip} and \texttt{sky}. Holding data transformations constant \textbf{(a)}, we posit that participants who focused on the \texttt{precip} attribute \textbf{(b)} versus the \texttt{sky} attribute \textbf{(c)} likely drew different conclusions, showing how attribute selection can influence insight discovery.\vspace{-2mm}}}
    \label{fig:case-studies:precip-sky}
\end{figure*}

\subsection{\revised{Use Case 3: Analyzing Participants' Insights}}
\label{sec:case-studies:battle-heer}

\revised{Although insight-based user studies have been critical to understanding how people form insights, participants' insights are generally self-reported, requiring a means of \emph{validating} insight quality. Traditionally this has been done by hand~\cite{saraiya_insight-based_2005,saraiya_insight-based_2006,north_toward_2006,liu_effects_2014,guo_case_2016}. However, recent considers how to partially automate the validation process~\cite{zgraggen_investigating_2018,battle_characterizing_2019,he_characterizing_2020}.}

\revised{\model\ provides a convenient structure for validating participants' insights and could facilitate further automation. We demonstrate this benefit by recreating insights reported by Battle and Heer from their study of how analysts explore data in Tableau~\cite{battle_characterizing_2019}.
We focus on task ``T3'' for the wildlife strikes dataset\footnote{\url{https://wildlife.faa.gov/search}}, which asks: \emph{``What relationships (if any) do you observe involving weather conditions and strike frequency, or counts over time?''} The task also specifies three attributes to consider: \texttt{precip}, \texttt{sky}, and \texttt{incident\_date}. We recreate two \emph{contradictory} answers observed by Battle and Heer: (A) strikes are not correlated with time (reported by 13 participants) and (B) bad weather leads to more strikes over time (reported by 3 participants). We use \model\ to shed light on the discrepancy among participants (see \autoref{fig:case-studies:precip-sky}.)}

\begin{enumerate}[itemsep=0em,topsep=0em,label=\Alph*)]
\item \revised{\emph{Analyze the \texttt{precip} Attribute.} First, we analyze the incidence of wildlife strikes using the \texttt{precip} attribute by applying a series of data transformations to: remove null \texttt{precip} entries on lines 4-7, extract the year from each \texttt{incident\_date} on line 11, and count the total incidents observed per year, grouped by \texttt{precip} conditions (e.g., ``fog,'' ``rain,'' etc.) on lines 15-26. Overall, we see that incidents do \emph{not} appear to increase with time, with the exception of ``rain'' conditions, shown in \autoref{fig:case-studies:precip}. We can record these findings in a new analytic knowledge node named \texttt{precipNode} in \autoref{fig:case-studies:birdstrikes-code}.}

\item \revised{\emph{Analyze the \texttt{sky} Attribute.} Second, we repeat this analysis, but replace \texttt{precip} with \texttt{sky} on lines 6 and 12 in our code, denoted in yellow in \autoref{fig:case-studies:birdstrikes-code}. In this case, we see a steady increase in incidents per year for all observed weather conditions, shown in \autoref{fig:case-studies:sky}. We can record these findings in a new analytic knowledge node named \texttt{skyNode} in \autoref{fig:case-studies:birdstrikes-code}.}
\end{enumerate}

\subsubsection{\revised{Benefits of The Formalism}}
\revised{We see that even when performing the same data transformations, participants could still derive drastically different answers based on which attributes were analyzed. These results suggest that even if participants' analyzed both attributes while completing T3, their answers were likely influenced by which attribute they favored, \texttt{precip} or \texttt{sky}. Since Battle and Heer focused on analyzing interaction sequences~\cite{battle_characterizing_2019}, they may have overlooked potential \emph{structural similarities} in participants' answers.
By using a consistent structure to represent analytic knowledge, we see how the data can influence which conclusions participants draw regardless of which interactions were performed.}

\vspace{-1mm}
\subsection{\revised{Use Case 4: Comparing Insight Studies}}
\label{sec:case-studies:comparison}
\vspace{-1mm}

\begin{figure}
    \centering
    \includegraphics[width=1.0\columnwidth]{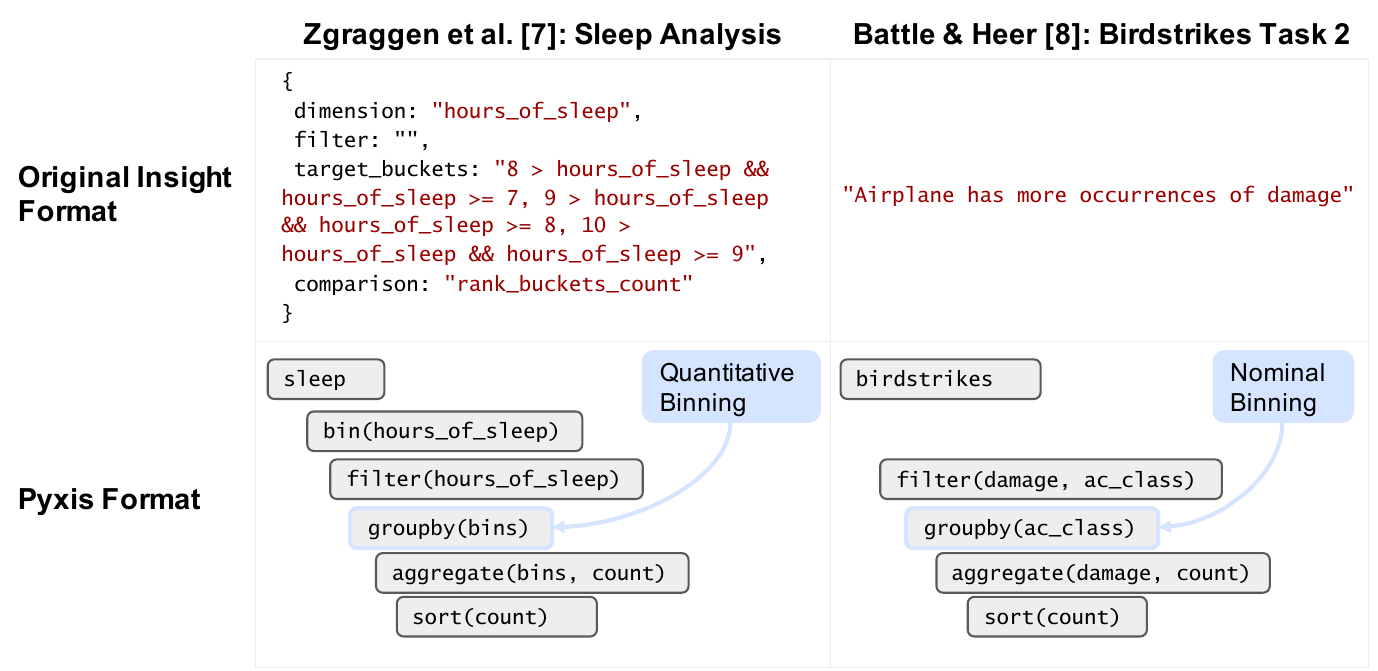}
    \vspace{-2mm}
    \caption{\revised{Although the analytic knowledge observed by Zgraggen et al.~\cite{zgraggen_investigating_2018} and Battle and Heer~\cite{battle_characterizing_2019} have very different formats, they share striking structural similarities when implemented using our formalism.}}
    \label{fig:zgraggen_v_battle}
\end{figure}

In this \revised{use case}, we seek to understand how existing insight-based studies define insight and \revised{application-driven} definitions compare with existing theories. In other words: \emph{how useful are existing theories for existing studies?}

\vspace{2mm}
\subsubsection{A Tale of Two Studies} For the sake of space, we limited our analysis to two insight-based studies by Zgraggen et al.~\cite{zgraggen_investigating_2018} and Battle and Heer~\cite{battle_characterizing_2019}. We used \model\ to implement all of the insights shared by Zgraggen et al. for their sleep study dataset exploration task, i.e., from Figure 3 of their paper~\cite{zgraggen_investigating_2018}. We also implemented three insights observed by Battle and Heer as analysts explored wildlife strikes data~\cite{battle_characterizing_2019}.
Examples of insights from both studies are shown in \autoref{fig:zgraggen_v_battle}, using the original formats at the top (lightly edited to maximize readability) and the corresponding \model\ data structures  (in this case, data transformations) at the bottom.

\vspace{2mm}
\subsubsection{These Studies Share Striking Similarities} Upon first impression, these insights may appear to be drastically different. However, we observed striking structural similarities. First, both studies \emph{record insights as analytic knowledge only} and omit domain knowledge. Second, the vast \emph{majority of analytic knowledge were data transformations only}, but some data relationships are also observed, e.g., linear correlations. Third, our implemented data structures were very similar between the two studies. For example, most data transformations involved filtering and/or grouping by an independent variable and aggregating a dependent variable to produce summary statistics. An example is shown in \autoref{fig:zgraggen_v_battle}, where Zgraggen et al. observe quantitative binning on the \texttt{hours\_of\_sleep} attribute and Battle and Heer observe nominal binning using the \texttt{ac\_class} attribute. Quantitative binning requires an extra step to discretize \texttt{hours\_of\_sleep} into bins, but otherwise, \emph{their data transformations have identical structures}.

\vspace{2mm}
\subsubsection{Existing Theory Only Covers a Subset of \revised{``Insights''}}
In some cases, we see strong overlaps between observed analytic knowledge from these studies and existing theory papers. For example, the \texttt{rank\_buckets\_count} insight in \autoref{fig:zgraggen_v_battle} bears a strong resemblance to North's rent distribution example~\cite{north_toward_2006} (see \autoref{sec:case-studies:complexity}).
However, North's examples only provide a partial fit for the analytic knowledge observed by Zgraggen et al.~\cite{zgraggen_investigating_2018} and Battle and Heer~\cite{battle_characterizing_2019}. For example, North does not provide explicit examples of multivariate relationships. Definitions emphasizing domain knowledge would also be a poor fit (e.g.,  \cite{gotz_interactive_2006}) since domain knowledge is not a focus of these studies.

Instead, theories that emphasize analytic knowledge as a whole are more appropriate. For example, Yang et al. formalize data facts, which encompass both data transformations and data relationships~\cite{yang_chen_toward_2009}. Kandogan and Engelke provide an alternative definition of analytic knowledge based primarily on relational query patterns~\cite{kandogan2018towards}.
Similarly, Demiralp et al. define insights as ``a strong manifestation of a distributional property of the data, such as strong correlation, tight clustering, low dispersion, and so on.''~\cite{demiralp_foresight_2017}.
However, neither paper cites these definitions.
We find that many insight-based studies observe the discovery of both data transformations and data relationships during visual analysis sessions~\cite{saraiya_insight-based_2005,smuc_score_2009,kandogan2018towards,he_characterizing_2020}. We also observe a recurring theme: these studies tend to reference the more popular definitions of insight rather than the most relevant ones. This pattern reveals an important limitation to existing theory: it is difficult to identify and apply the most relevant theories to insight-based studies. Our formalism takes a step towards addressing these challenges by providing a framework for navigating existing theories of insight.

\vspace{2mm}
\subsubsection{\revised{Benefits of the Formalism}} Given the importance of data distributions as analytic knowledge, we could reuse the corresponding \model{} objects as \emph{templates} for extracting this knowledge from new datasets, e.g., for visualization recommendation~\cite{zeng2022evaluation}. Further, rather than limiting recommendations to quantitative binning (i.e., standard histograms), we can also apply qualitative binning as observed by Battle and Heer~\cite{battle_characterizing_2019}. To use \model{} objects as templates, we can replace the assigned attributes in the \model{} objects (e.g., \texttt{hours\_of\_sleep}, \texttt{ac\_class}) with equivalent attribute(s) from a target dataset. This approach also reveals the potential of \model\ as a language for \emph{taxonomizing} \revised{observed insights}, \revised{thereby} extending existing categorizations (\autoref{sec:background:insight:categories}) with a \revised{corpus of exemplars amenable to quantitative meta-analysis}.

\section{Discussion and Future Work}
\label{sec:discussion}
\vspace{-1mm}

Researchers largely agree on the \emph{structure} of insights, i.e., their major building blocks, but not their \emph{semantics}, i.e., how these building blocks are interpreted.
We hone existing structural consistencies into a unified formalism to quantify the complexity and scope of observed knowledge.
We highlight exciting future directions based on this work.

\vspace{-1mm}
\subsection{Declarative Specification of \revised{Insights}}
\vspace{-1mm}
We observe a consistent progression from an unstructured recording of analytic knowledge towards \emph{declarative specification} of analytic knowledge.
This is exemplified in the progressions of building blocks proposed by North~\cite{north_comparison_2011}, Kandogan and Engelke~\cite{kandogan2018towards}, Zgraggen et al.~\cite{zgraggen_investigating_2018}, and Andrienko and Andrienko~\cite{andrienko2006exploratory} as well as in related work by Suh et al.~\cite{suh2022grammar}.
These works converge on expressing a user’s interpretation of data in terms of \emph{relational queries} rather than a series of logged system interactions or interface manipulations. An example from Zgraggen et al. is shown in \autoref{fig:zgraggen_v_battle}.
This direction can also provide interesting opportunities to intersect with related areas, such as databases and programming languages research~\cite{battle2021structured}.
Still, there are clear limitations to this approach revealed by our formalism (see \autoref{sec:case-studies:complexity}). For example, relational query languages like SQL lack declarative representations for multivariate and univariate data relationships. We highlight this as a critical gap to be filled in future theories.


\vspace{-1mm}
\subsection{Supporting Domain Knowledge}
\vspace{-1mm}

We observed a recurring theme across insight-based theories and studies: they lack precise specifications for domain knowledge, hindering our community's ability to reason about user knowledge as a whole.
For example, a correlation between two generic variables is meaningless without domain context, e.g., a correlation between certain drugs and patient health outcomes~\cite{he_characterizing_2020} or the incidence of crime and racial injustice in Baltimore~\cite{mathisen_insideinsights_2019}. 
This is not solely a visualization provenance issue. For example, recent research shows how \revised{visualization} recommendation systems~\cite{zehrungsinghal2021vis,zeng2022evaluation} and visualization languages~\cite{battle2022exploring} can be hindered by their inability to incorporate domain and task context. Thus, incorporating \revised{user domain} knowledge appears to be a consistent challenge within visualization research.

As a first step, \textbf{we observe a need for more precise methods for expressing user domain knowledge}, ideally at the theoretical level. 
In this way, we can capture user domain knowledge with equivalent precision as analytic knowledge, allowing our community to more accurately model user learning and knowledge building over time.

One possibility for refining representations of domain knowledge is to exploit \revised{formalisms from} knowledge graphs, which could enable us to automate traditionally manual practices using graph-based algorithms, e.g., to quantify the depth and breadth of user knowledge (\autoref{sec:case-studies:complexity}), to detect patterns in knowledge acquisition~\cite{toniolo2023human,kandogan2018towards}, or even to provide additional context to visualization tools such as for visualization recommendation tasks~\cite{li2022kg4vis}. Further, knowledge graphs could support bias detection by detecting problematic links between domain and analytic knowledge, and analytic knowledge and conclusions that are drawn.

\vspace{-1mm}
\subsection{Sharing Insight Data}
\vspace{-1mm}
A critical second step is to \textbf{prioritize sharing of insight data.}
For example, although Liu and Heer share their annotations for user insights, it is impossible to extract the specific data users were analyzing in imMens in tandem with these annotations~\cite{liu_effects_2014}.
Zgraggen et al.~\cite{zgraggen_investigating_2018} and Kandogan and Engelke~\cite{kandogan2018towards} derive insight specifications from their study data but only share a handful of examples as brief anecdotes in their papers.
Similarly, Guo et al. only describe a few example insights from their study in their paper~\cite{guo_case_2016}.
In contrast, best practices in visualization provenance encourage consistent tracking, sharing, and reuse of recorded system interaction logs (e.g., \cite{gathani2022grammar,feng_patterns_2019,battle_characterizing_2019,cutler2020trrack}). 

Insight \revised{corpora} could open new avenues for visualization research. For example, recorded insights could be aggregated to form reusable insight \emph{templates} to \revised{detect} insights within \revised{new} datasets. Insight datasets could also be used as unit tests for evaluating insight specifications, for example, by adapting the coverage and diversity measures proposed by Gathani et al. to apply to insight specifications~\cite{gathani2022grammar}.
Our formalism eases this burden somewhat by providing precise specifications for the core building blocks of insights. In this way, researchers can choose which building blocks to focus on and use \model\ to share observed instances of these building blocks from their user studies. 


\section{Conclusion}
\label{sec:conclusion}

Reviewing the literature, we find that researchers seem to agree on the \emph{structure} of visual analysis insights, i.e., their major building blocks, but not the \emph{semantics} of insights, i.e., how these building blocks are interpreted.
We propose a unified formalism that integrates multiple theoretical definitions of insight and contribute a toolkit called \model\ for specifying insights using our formalism.
We \revised{demonstrate how to} use \model\ to implement existing definitions of insight and compare the resulting structures. We find that current definitions fail to support rich specifications of domain knowledge and analytic knowledge, revealing exciting visualization research opportunities in domain knowledge representation, automated analytical reasoning, and collaborations with data management and data mining researchers.

\section*{Acknowledgements}
This research was supported by grants \#2141506, \#2142977, and \#2118201 from the National Science Foundation (NSF).

\bibliographystyle{IEEEtran}
\bibliography{IEEEabrv,vis-task-theory2.bib}

\begin{thebibliography}{10}
\providecommand{\url}[1]{#1}
\csname url@samestyle\endcsname
\providecommand{\newblock}{\relax}
\providecommand{\bibinfo}[2]{#2}
\providecommand{\BIBentrySTDinterwordspacing}{\spaceskip=0pt\relax}
\providecommand{\BIBentryALTinterwordstretchfactor}{4}
\providecommand{\BIBentryALTinterwordspacing}{\spaceskip=\fontdimen2\font plus
\BIBentryALTinterwordstretchfactor\fontdimen3\font minus
  \fontdimen4\font\relax}
\providecommand{\BIBforeignlanguage}[2]{{%
\expandafter\ifx\csname l@#1\endcsname\relax
\typeout{** WARNING: IEEEtran.bst: No hyphenation pattern has been}%
\typeout{** loaded for the language `#1'. Using the pattern for}%
\typeout{** the default language instead.}%
\else
\language=\csname l@#1\endcsname
\fi
#2}}
\providecommand{\BIBdecl}{\relax}
\BIBdecl

\bibitem{hullman2019purpose}
J.~Hullman, ``The purpose of visualization is insight, not pictures: An
  interview with ben shneiderman,'' \emph{ACM Interactions}, 2019.

\bibitem{battle2016dynamic}
L.~Battle, R.~Chang, and M.~Stonebraker, ``Dynamic {Prefetching} of {Data}
  {Tiles} for {Interactive} {Visualization},'' ser. {SIGMOD} '16.\hskip 1em
  plus 0.5em minus 0.4em\relax ACM, 2016, pp. 1363--1375.

\bibitem{monadjemi_competing_2020}
S.~Monadjemi, R.~Garnett, and A.~Ottley, ``Competing {Models}: {Inferring}
  {Exploration} {Patterns} and {Information} {Relevance} via {Bayesian} {Model}
  {Selection},'' \emph{IEEE TVCG}, pp. 1--1, 2020.

\bibitem{gathani2022grammar}
S.~Gathani, S.~Monadjemi, A.~Ottley, and L.~Battle, ``A grammar-based approach
  for applying visualization taxonomies to interaction logs,'' in \emph{CGF},
  vol.~41, no.~3.\hskip 1em plus 0.5em minus 0.4em\relax Wiley Online Library,
  2022, pp. 489--500.

\bibitem{pohl2012analysing}
M.~Pohl, S.~Wiltner, S.~Miksch, W.~Aigner, and A.~Rind, ``Analysing
  interactivity in information visualisation,'' \emph{KI-K{\"u}nstliche
  Intelligenz}, vol.~26, pp. 151--159, 2012.

\bibitem{north_toward_2006}
C.~North, ``Toward measuring visualization insight,'' \emph{IEEE CG\&A},
  vol.~26, no.~3, pp. 6--9, May 2006.

\bibitem{zgraggen_investigating_2018}
E.~Zgraggen, Z.~Zhao, R.~Zeleznik, and T.~Kraska, ``Investigating the {Effect}
  of the {Multiple} {Comparisons} {Problem} in {Visual} {Analysis},'' ser.
  {CHI} '18.\hskip 1em plus 0.5em minus 0.4em\relax ACM, Apr. 2018, pp. 1--12.

\bibitem{battle_characterizing_2019}
L.~Battle and J.~Heer, ``\BIBforeignlanguage{en}{Characterizing {Exploratory}
  {Visual} {Analysis}: {A} {Literature} {Review} and {Evaluation} of {Analytic}
  {Provenance} in {Tableau}},'' \emph{\BIBforeignlanguage{en}{CGF}}, vol.~38,
  no.~3, pp. 145--159, 2019.

\bibitem{rind_task_2016}
A.~Rind, W.~Aigner, M.~Wagner, S.~Miksch, and T.~Lammarsch,
  ``\BIBforeignlanguage{en}{Task {Cube}: {A} three-dimensional conceptual space
  of user tasks in visualization design and evaluation},''
  \emph{\BIBforeignlanguage{en}{Information Visualization}}, vol.~15, no.~4,
  pp. 288--300, Oct. 2016, publisher: SAGE Publications.

\bibitem{satyanarayan2017vegalite}
A.~Satyanarayan, D.~Moritz, K.~Wongsuphasawat, and J.~Heer, ``Vega-lite: A
  grammar of interactive graphics,'' \emph{IEEE TVCG}, vol.~23, no.~1, pp.
  341--350, 2017.

\bibitem{McNutt2021Templates}
A.~M. McNutt and R.~Chugh, \emph{Integrated Visualization Editing via
  Parameterized Declarative Templates}, ser. CHI '21.\hskip 1em plus 0.5em
  minus 0.4em\relax ACM, 2021.

\bibitem{battle2023exactly}
L.~Battle and A.~Ottley, ``What exactly is an insight? a literature review,''
  \emph{arXiv preprint arXiv:2307.06551 (to appear in IEEE VIS 2023 -- Short
  Papers)}, 2023.

\bibitem{north_comparison_2011}
C.~North, P.~Saraiya, and K.~Duca, ``\BIBforeignlanguage{en}{A comparison of
  benchmark task and insight evaluation methods for information
  visualization},'' \emph{\BIBforeignlanguage{en}{Information Visualization}},
  vol.~10, no.~3, pp. 162--181, Jul. 2011.

\bibitem{chang_defining_2009}
R.~Chang, C.~Ziemkiewicz, T.~M. Green, and W.~Ribarsky, ``Defining {Insight}
  for {Visual} {Analytics},'' \emph{IEEE CG\&A}, vol.~29, no.~2, pp. 14--17,
  Mar. 2009.

\bibitem{saraiya_insight-based_2005}
P.~Saraiya, C.~North, and K.~Duca, ``\BIBforeignlanguage{en}{An
  {Insight}-{Based} {Methodology} for {Evaluating} {Bioinformatics}
  {Visualizations}},'' \emph{\BIBforeignlanguage{en}{IEEE TVCG}}, vol.~11,
  no.~4, pp. 443--456, Jul. 2005.

\bibitem{saraiya_evaluation_2004}
------, ``An {Evaluation} of {Microarray} {Visualization} {Tools} for
  {Biological} {Insight},'' in \emph{{IEEE} {Symposium} on {Information}
  {Visualization}}, Oct. 2004, pp. 1--8, iSSN: 1522-404X.

\bibitem{choe_characterizing_2015}
E.~K. Choe, B.~Lee, and m.~c. schraefel, ``Characterizing {Visualization}
  {Insights} from {Quantified} {Selfers}' {Personal} {Data} {Presentations},''
  \emph{IEEE CG\&A}, vol.~35, no.~4, pp. 28--37, Jul. 2015.

\bibitem{smuc_score_2009}
M.~Smuc, E.~Mayr, T.~Lammarsch, W.~Aigner, S.~Miksch, and J.~Gärtner, ``To
  {Score} or {Not} to {Score}? {Tripling} {Insights} for {Participatory}
  {Design},'' \emph{IEEE CG\&A}, vol.~29, no.~3, pp. 29--38, May 2009.

\bibitem{gotz_interactive_2006}
D.~Gotz, M.~X. Zhou, and V.~Aggarwal, ``Interactive {Visual} {Synthesis} of
  {Analytic} {Knowledge},'' in \emph{2006 {IEEE} {Symposium} {On} {Visual}
  {Analytics} {Science} {And} {Technology}}, Oct. 2006, pp. 51--58.

\bibitem{pousman_casual_2007}
Z.~Pousman, J.~Stasko, and M.~Mateas, ``Casual {Information} {Visualization}:
  {Depictions} of {Data} in {Everyday} {Life},'' \emph{IEEE TVCG}, vol.~13,
  no.~6, pp. 1145--1152, Nov. 2007.

\bibitem{liu_effects_2014}
Z.~Liu and J.~Heer, ``The {Effects} of {Interactive} {Latency} on {Exploratory}
  {Visual} {Analysis},'' \emph{IEEE TVCG}, vol.~20, no.~12, pp. 2122--2131,
  Dec. 2014.

\bibitem{karer2021insight}
B.~Karer, H.~Hagen, and D.~J. Lehmann, ``Insight beyond numbers: The impact of
  qualitative factors on visual data analysis,'' \emph{IEEE Transactions on
  Visualization and Computer Graphics}, vol.~27, no.~2, pp. 1011--1021, 2021.

\bibitem{zgraggen2017progressive}
E.~Zgraggen, A.~Galakatos, A.~Crotty, J.-D. Fekete, and T.~Kraska, ``How
  progressive visualizations affect exploratory analysis,'' \emph{IEEE TVCG},
  vol.~23, no.~8, pp. 1977--1987, 2017.

\bibitem{saraiya_insight-based_2006}
P.~Saraiya, C.~North, {Vy Lam}, and K.~Duca, ``An {Insight}-{Based}
  {Longitudinal} {Study} of {Visual} {Analytics},'' \emph{IEEE TVCG}, vol.~12,
  no.~6, pp. 1511--1522, Nov. 2006.

\bibitem{plaisant_promoting_2008}
C.~Plaisant, J.-D. Fekete, and G.~Grinstein, ``Promoting {Insight}-{Based}
  {Evaluation} of {Visualizations}: {From} {Contest} to {Benchmark}
  {Repository},'' \emph{IEEE TVCG}, vol.~14, no.~1, pp. 120--134, Jan. 2008.

\bibitem{yang_chen_toward_2009}
{Yang Chen}, {Jing Yang}, and W.~Ribarsky, ``Toward effective insight
  management in visual analytics systems,'' in \emph{2009 {IEEE} {Pacific}
  {Visualization} {Symposium}}, Apr. 2009, pp. 49--56, iSSN: 2165-8773.

\bibitem{sacha_knowledge_2014}
D.~Sacha, A.~Stoffel, F.~Stoffel, B.~C. Kwon, G.~Ellis, and D.~A. Keim,
  ``Knowledge {Generation} {Model} for {Visual} {Analytics},'' \emph{IEEE
  TVCG}, vol.~20, no.~12, pp. 1604--1613, Dec. 2014.

\bibitem{gomez_insight-_2014}
S.~R. Gomez, H.~Guo, C.~Ziemkiewicz, and D.~H. Laidlaw, ``An insight- and
  task-based methodology for evaluating spatiotemporal visual analytics,'' in
  \emph{2014 {IEEE} {Conference} on {Visual} {Analytics} {Science} and
  {Technology} ({VAST})}, Oct. 2014, pp. 63--72.

\bibitem{guo_case_2016}
H.~Guo, S.~R. Gomez, C.~Ziemkiewicz, and D.~H. Laidlaw, ``A {Case} {Study}
  {Using} {Visualization} {Interaction} {Logs} and {Insight} {Metrics} to
  {Understand} {How} {Analysts} {Arrive} at {Insights},'' \emph{IEEE TVCG},
  vol.~22, no.~1, pp. 51--60, Jan. 2016.

\bibitem{srinivasan_augmenting_2019}
A.~Srinivasan, S.~M. Drucker, A.~Endert, and J.~Stasko, ``Augmenting
  {Visualizations} with {Interactive} {Data} {Facts} to {Facilitate}
  {Interpretation} and {Communication},'' \emph{IEEE TVCG}, vol.~25, no.~1, pp.
  672--681, Jan. 2019.

\bibitem{kandogan2018towards}
E.~Kandogan and U.~Engelke, ``Towards a unified representation of insight in
  human-in-the-loop analytics: A user study,'' ser. HILDA'18.\hskip 1em plus
  0.5em minus 0.4em\relax ACM, 2018.

\bibitem{yi_understanding_2008}
J.~S. Yi, Y.-a. Kang, J.~T. Stasko, and J.~A. Jacko,
  ``\BIBforeignlanguage{en}{Understanding and characterizing insights: how do
  people gain insights using information visualization?}'' in
  \emph{\BIBforeignlanguage{en}{Proceedings of the 2008 conference on {BEyond}
  time and errors novel {evaLuation} methods for {Information} {Visualization}
  - {BELIV} '08}}.\hskip 1em plus 0.5em minus 0.4em\relax Florence, Italy: ACM
  Press, 2008, p.~1.

\bibitem{amar_knowledge_2005}
R.~A. Amar and J.~T. Stasko, ``Knowledge precepts for design and evaluation of
  information visualizations,'' \emph{IEEE TVCG}, vol.~11, no.~4, pp. 432--442,
  Jul. 2005.

\bibitem{shrinivasan_connecting_2009}
Y.~B. Shrinivasan, D.~Gotz, and J.~Lu, ``Connecting the dots in visual
  analysis,'' in \emph{{IEEE} VAST}, Oct. 2009, pp. 123--130.

\bibitem{shrinivasan_supporting_2008}
Y.~B. Shrinivasan and J.~J. van Wijk, ``Supporting the analytical reasoning
  process in information visualization,'' ser. {CHI} '08.\hskip 1em plus 0.5em
  minus 0.4em\relax ACM, Apr. 2008, pp. 1237--1246.

\bibitem{gotz_characterizing_2009}
D.~Gotz and M.~X. Zhou, ``\BIBforeignlanguage{en}{Characterizing {Users}'
  {Visual} {Analytic} {Activity} for {Insight} {Provenance}},''
  \emph{\BIBforeignlanguage{en}{Information Visualization}}, vol.~8, no.~1, pp.
  42--55, Jan. 2009, publisher: SAGE Publications.

\bibitem{he_characterizing_2020}
C.~He, L.~Micallef, L.~He, G.~Peddinti, T.~Aittokallio, and G.~Jacucci,
  ``Characterizing the {Quality} of {Insight} by {Interactions}: {A} {Case}
  {Study},'' \emph{IEEE TVCG}, pp. 1--1, 2020.

\bibitem{green_visual_2008}
T.~M. Green, W.~Ribarsky, and B.~Fisher, ``Visual analytics for complex
  concepts using a human cognition model,'' in \emph{{IEEE} {VAST}}, Oct. 2008,
  pp. 91--98.

\bibitem{willett2011commentspace}
W.~Willett, J.~Heer, J.~Hellerstein, and M.~Agrawala, ``Commentspace:
  Structured support for collaborative visual analysis,'' in \emph{Proceedings
  of the SIGCHI Conference on Human Factors in Computing Systems}, ser. CHI
  '11.\hskip 1em plus 0.5em minus 0.4em\relax ACM, 2011, p. 3131–3140.

\bibitem{dou_recovering_2009}
W.~Dou, D.~H. Jeong, F.~Stukes, W.~Ribarsky, H.~R. Lipford, and R.~Chang,
  ``Recovering {Reasoning} {Processes} from {User} {Interactions},'' \emph{IEEE
  CG\&A}, vol.~29, no.~3, pp. 52--61, May 2009.

\bibitem{mathisen_insideinsights_2019}
A.~Mathisen, T.~Horak, C.~N. Klokmose, K.~Grønbæk, and N.~Elmqvist,
  ``\BIBforeignlanguage{en}{{InsideInsights}: {Integrating} {Data}-{Driven}
  {Reporting} in {Collaborative} {Visual} {Analytics}},''
  \emph{\BIBforeignlanguage{en}{CGF}}, vol.~38, no.~3, pp. 649--661, 2019.

\bibitem{pike_science_2009}
W.~A. Pike, J.~Stasko, R.~Chang, and T.~A. O'Connell,
  ``\BIBforeignlanguage{en}{The {Science} of {Interaction}},''
  \emph{\BIBforeignlanguage{en}{Information Visualization}}, vol.~8, no.~4, pp.
  263--274, Jan. 2009, publisher: SAGE Publications.

\bibitem{toniolo2023human}
\BIBentryALTinterwordspacing
A.~Toniolo, F.~Cerutti, T.~J. Norman, N.~Oren, J.~A. Allen, M.~Srivastava, and
  P.~Sullivan, ``Human-machine collaboration in intelligence analysis: An
  expert evaluation,'' \emph{Intelligent Systems with Applications}, vol.~17,
  p. 200151, 2023. [Online]. Available:
  \url{https://www.sciencedirect.com/science/article/pii/S2667305322000886}
\BIBentrySTDinterwordspacing

\bibitem{brehmer_multi-level_2013}
M.~Brehmer and T.~Munzner, ``A {Multi}-{Level} {Typology} of {Abstract}
  {Visualization} {Tasks},'' \emph{IEEE TVCG}, vol.~19, no.~12, pp. 2376--2385,
  Dec. 2013.

\bibitem{lam_bridging_2018}
H.~Lam, M.~Tory, and T.~Munzner, ``Bridging from {Goals} to {Tasks} with
  {Design} {Study} {Analysis} {Reports},'' \emph{IEEE TVCG}, vol.~24, no.~1,
  pp. 435--445, Jan. 2018.

\bibitem{amar_low-level_2005}
R.~Amar, J.~Eagan, and J.~Stasko, ``Low-level components of analytic activity
  in information visualization,'' in \emph{{INFOVIS} 2005.}, Oct. 2005, pp.
  111--117, iSSN: 1522-404X.

\bibitem{kang_examining_2012}
Y.~Kang and J.~Stasko, ``Examining the {Use} of a {Visual} {Analytics} {System}
  for {Sensemaking} {Tasks}: {Case} {Studies} with {Domain} {Experts},''
  \emph{IEEE TVCG}, vol.~18, no.~12, pp. 2869--2878, Dec. 2012.

\bibitem{bertin1983semiology}
J.~Bertin, \emph{Semiology of graphics}.\hskip 1em plus 0.5em minus 0.4em\relax
  University of Wisconsin press, 1983.

\bibitem{andrienko2006exploratory}
N.~Andrienko and G.~Andrienko, \emph{Exploratory analysis of spatial and
  temporal data: a systematic approach}.\hskip 1em plus 0.5em minus 0.4em\relax
  Springer Science \& Business Media, 2006.

\bibitem{codd1970relational}
E.~F. Codd, ``A relational model of data for large shared data banks,''
  \emph{Commun. ACM}, vol.~13, no.~6, p. 377–387, jun 1970.

\bibitem{wilkinson2012grammar}
L.~Wilkinson, ``The grammar of graphics,'' in \emph{Handbook of computational
  statistics}.\hskip 1em plus 0.5em minus 0.4em\relax Springer, 2012, pp.
  375--414.

\bibitem{yun2021knowledge}
W.~Yun, X.~Zhang, Z.~Li, H.~Liu, and M.~Han, ``Knowledge modeling: A survey of
  processes and techniques,'' \emph{International Journal of Intelligent
  Systems}, vol.~36, no.~4, pp. 1686--1720, 2021.

\bibitem{kale2022causal}
A.~Kale, Y.~Wu, and J.~Hullman, ``Causal support: Modeling causal inferences
  with visualizations,'' \emph{IEEE TVCG}, vol.~28, no.~1, pp. 1150--1160,
  2022.

\bibitem{xu_survey_2020}
K.~Xu, A.~Ottley, C.~Walchshofer, M.~Streit, R.~Chang, and J.~Wenskovitch,
  ``\BIBforeignlanguage{en}{Survey on the {Analysis} of {User} {Interactions}
  and {Visualization} {Provenance}},'' \emph{\BIBforeignlanguage{en}{CGF}},
  vol.~39, no.~3, pp. 757--783, 2020.

\bibitem{hogan2021knowledge}
A.~Hogan, E.~Blomqvist, M.~Cochez, C.~D’amato, G.~D. Melo, C.~Gutierrez,
  S.~Kirrane, J.~E.~L. Gayo, R.~Navigli, S.~Neumaier, A.-C.~N. Ngomo,
  A.~Polleres, S.~M. Rashid, A.~Rula, L.~Schmelzeisen, J.~Sequeda, S.~Staab,
  and A.~Zimmermann, ``Knowledge graphs,'' \emph{ACM Comput. Surv.}, vol.~54,
  no.~4, jul 2021.

\bibitem{ji2021survey}
S.~Ji, S.~Pan, E.~Cambria, P.~Marttinen, and S.~Y. Philip, ``A survey on
  knowledge graphs: Representation, acquisition, and applications,'' \emph{IEEE
  Transactions on Neural Networks and Learning Systems}, vol.~33, no.~2, pp.
  494--514, 2021.

\bibitem{bizer2009linked}
C.~Bizer, T.~Heath, T.~Berners-Lee \emph{et~al.}, ``Linked data-the story so
  far,'' \emph{IJSWIS}, vol.~5, no.~3, pp. 1--22, 2009.

\bibitem{nonaka2007knowledge}
I.~Nonaka and H.~Takeuchi, ``The knowledge-creating company,'' \emph{Harvard
  business review}, vol.~85, no. 7/8, p. 162, 2007.

\bibitem{andrienko_viewing_2018}
N.~Andrienko, T.~Lammarsch, G.~Andrienko, G.~Fuchs, D.~Keim, S.~Miksch, and
  A.~Rind, ``\BIBforeignlanguage{en}{Viewing {Visual} {Analytics} as {Model}
  {Building}},'' \emph{\BIBforeignlanguage{en}{CGF}}, vol.~37, no.~6, pp.
  275--299, 2018.

\bibitem{zehrungsinghal2021vis}
R.~Zehrung, A.~Singhal, M.~Correll, and L.~Battle, ``Vis ex machina: An
  analysis of trust in human versus algorithmically generated visualization
  recommendations,'' ser. CHI '21.\hskip 1em plus 0.5em minus 0.4em\relax ACM,
  2021.

\bibitem{harrison2014ranking}
L.~Harrison, F.~Yang, S.~Franconeri, and R.~Chang, ``Ranking visualizations of
  correlation using weber's law,'' \emph{IEEE TVCG}, vol.~20, no.~12, pp.
  1943--1952, 2014.

\bibitem{demiralp_foresight_2017}
{Demiralp, Çağatay and Haas, Peter J. and Parthasarathy, Srinivasan and
  Pedapati, Tejaswini}, ``Foresight: recommending visual insights,''
  \emph{Proc. VLDB Endow.}, vol.~10, no.~12, pp. 1937--1940, Aug. 2017.

\bibitem{ottley_follow_2019}
A.~Ottley, R.~Garnett, and R.~Wan, ``\BIBforeignlanguage{en}{Follow {The}
  {Clicks}: {Learning} and {Anticipating} {Mouse} {Interactions} {During}
  {Exploratory} {Data} {Analysis}},'' \emph{\BIBforeignlanguage{en}{CGF}},
  vol.~38, no.~3, pp. 41--52, 2019.

\bibitem{stolte2002polaris}
C.~Stolte, D.~Tang, and P.~Hanrahan, ``Polaris: a system for query, analysis,
  and visualization of multidimensional relational databases,'' \emph{IEEE
  TVCG}, vol.~8, no.~1, pp. 52--65, 2002.

\bibitem{shneiderman_eyes_1996}
B.~Shneiderman, ``The eyes have it: a task by data type taxonomy for
  information visualizations,'' in \emph{Proceedings 1996 {IEEE} {Symposium} on
  {Visual} {Languages}}, Sep. 1996, pp. 336--343, iSSN: 1049-2615.

\bibitem{meijer2011world}
E.~Meijer, ``The world according to linq,'' \emph{CACM}, vol.~54, no.~10, pp.
  45--51, 2011.

\bibitem{chandramouli2014trill}
B.~Chandramouli, J.~Goldstein, M.~Barnett, R.~DeLine, D.~Fisher, J.~C. Platt,
  J.~F. Terwilliger, and J.~Wernsing, ``Trill: A high-performance incremental
  query processor for diverse analytics,'' \emph{Proc. VLDB Endow.}, vol.~8,
  no.~4, p. 401–412, dec 2014.

\bibitem{pirolli_sensemaking_2005}
P.~Pirolli and S.~Card, ``The sensemaking process and leverage points for
  analyst technology as identified through cognitive task analysis,'' in
  \emph{Proceedings of {International} {Conference} on {Intelligence}
  {Analysis}}, vol.~5, Jan. 2005.

\bibitem{vega2023transforms}
Vega{ }Development{ }Team, ``Transforms | vega,''
  \url{https://vega.github.io/vega/docs/transforms/}, 2023.

\bibitem{heer2021arquero}
J.~Heer, ``Arquero | arquero,'' \url{https://uwdata.github.io/arquero/}, 2021.

\bibitem{jain2016sqlshare}
S.~Jain, D.~Moritz, D.~Halperin, B.~Howe, and E.~Lazowska, ``Sqlshare: Results
  from a multi-year sql-as-a-service experiment,'' ser. SIGMOD '16.\hskip 1em
  plus 0.5em minus 0.4em\relax ACM, 2016, p. 281–293.

\bibitem{mccabe1976complexity}
T.~McCabe, ``A complexity measure,'' \emph{IEEE Transactions on Software
  Engineering}, vol. SE-2, no.~4, pp. 308--320, 1976.

\bibitem{zeng2022evaluation}
Z.~Zeng, P.~Moh, F.~Du, J.~Hoffswell, T.~Y. Lee, S.~Malik, E.~Koh, and
  L.~Battle, ``An evaluation-focused framework for visualization recommendation
  algorithms,'' \emph{IEEE TVCG}, vol.~28, no.~1, pp. 346--356, 2022.

\bibitem{suh2022grammar}
A.~Suh, Y.~Jiang, A.~Mosca, E.~Wu, and R.~Chang, ``A grammar for
  hypothesis-driven visual analysis,'' \emph{arXiv preprint arXiv:2204.14267},
  2022.

\bibitem{battle2021structured}
L.~Battle and C.~Scheidegger, ``A structured review of data management
  technology for interactive visualization and analysis,'' \emph{IEEE TVCG},
  vol.~27, no.~2, pp. 1128--1138, 2021.

\bibitem{battle2022exploring}
L.~Battle, D.~Feng, and K.~Webber, ``Exploring d3 implementation challenges on
  stack overflow,'' in \emph{IEEE VIS}, 2022, pp. 1--5.

\bibitem{li2022kg4vis}
H.~Li, Y.~Wang, S.~Zhang, Y.~Song, and H.~Qu, ``Kg4vis: A knowledge graph-based
  approach for visualization recommendation,'' \emph{IEEE Transactions on
  Visualization and Computer Graphics}, vol.~28, no.~1, pp. 195--205, 2022.

\bibitem{feng_patterns_2019}
M.~Feng, E.~Peck, and L.~Harrison, ``Patterns and {Pace}: {Quantifying}
  {Diverse} {Exploration} {Behavior} with {Visualizations} on the {Web},''
  \emph{IEEE TVCG}, vol.~25, no.~1, pp. 501--511, Jan. 2019.

\bibitem{cutler2020trrack}
Z.~Cutler, K.~Gadhave, and A.~Lex, ``Trrack: A library for provenance-tracking
  in web-based visualizations,'' in \emph{IEEE VIS}, 2020, pp. 116--120.

\end{thebibliography}


\vspace{-1in}

\begin{IEEEbiography}[{\includegraphics[width=1in,height=1.25in,clip,keepaspectratio]{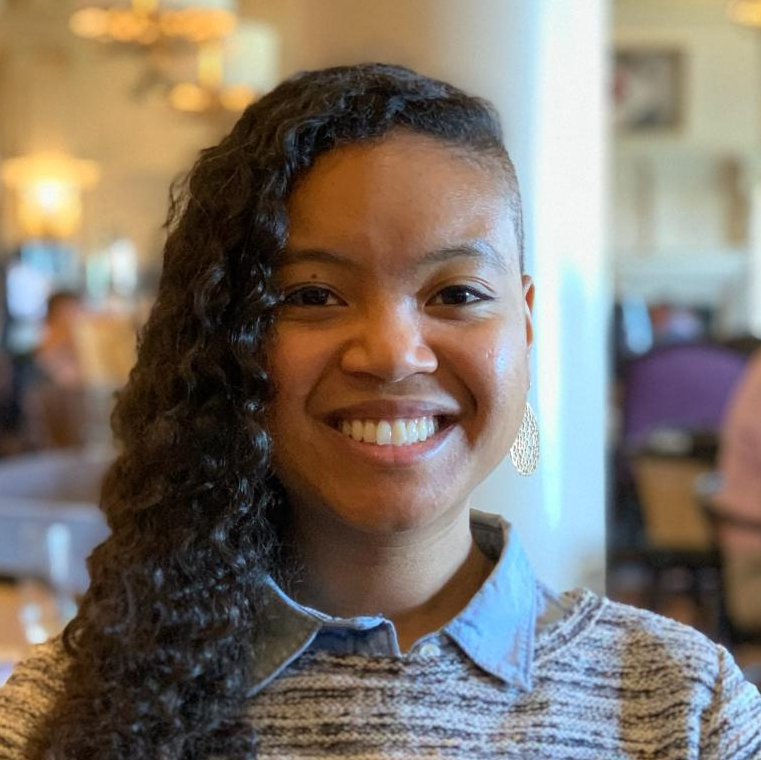}}]{Leilani Battle} is an Assistant Professor in the Paul G. Allen School for Computer Science \& Engineering at the University of Washington. Her research focuses on developing interactive data-intensive systems that aid analysts in performing complex data exploration and analysis. She completed a postdoc in the UW IDL and earned a PhD degree from MIT in the MIT Database Group.
\end{IEEEbiography}

\vspace{-1in}

\begin{IEEEbiography}[{\includegraphics[width=1in,height=1.25in,clip,keepaspectratio]{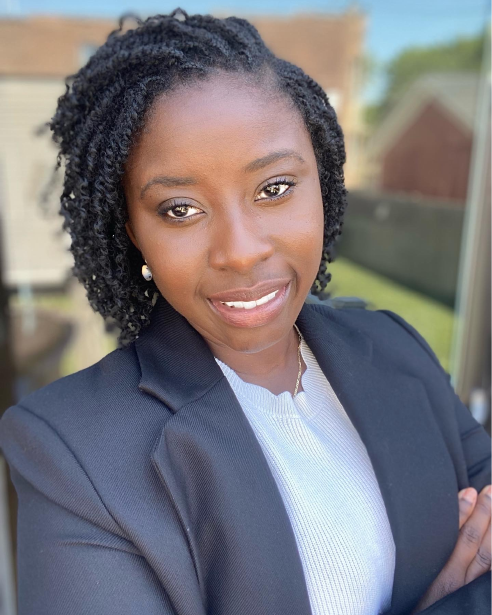}}]
{Alvitta Ottley} is an Associate Professor in Computer Science \& Engineering Department at Washington University in St. Louis, Missouri, USA.
Her research uses interdisciplinary approaches to solve problems such as how best to display information for effective decision-making and how to design human-in-the-loop visual analytics interfaces that are more attuned to how people think. She earned a PhD degree from Tufts University in the VALT group.
\end{IEEEbiography}


\end{document}